\documentclass[reprint,nofootinbib,amsmath,amssymb,floatfix]{revtex4-2}
\usepackage{graphicx}
\usepackage{dcolumn}
\usepackage[multiple]{footmisc}
\usepackage{appendix}
\usepackage{color}
\usepackage{multirow}
\usepackage{url}
\usepackage{amsmath}
\usepackage{amsfonts}
\usepackage[colorlinks=true, allcolors=blue]{hyperref}
\usepackage{caption}
\usepackage{subcaption}
\usepackage{verbatim}
\usepackage{amssymb}
\usepackage{hyperref}
\usepackage{enumitem}
\usepackage{float} % 插入图片的宏包

\newcommand{\bea}{\begin{eqnarray}}
\newcommand{\eea}{\end{eqnarray}}
\newcommand{\nn}{\nonumber}

\begin{document}

%填写标题作者单位等信息
\title{Radii of spherical timelike geodesics in Kerr-Newman black holes}
\author{Wei Huang\textsuperscript{1,2}}
\author{Jun-Xu Chen\textsuperscript{3,4}}
\author{Jia-Hui Huang\textsuperscript{1,2}} 
\email{huangjh@m.scnu.edu.cn}
\date{\today}

\affiliation {$^{1}$ Key Laboratory of Atomic and Subatomic Structure and Quantum Control (Ministry of Education), Guangdong Basic Research Center of Excellence for Structure and Fundamental Interactions of Matter, School of Physics, South China Normal University, Guangzhou 510006, China}
\affiliation {$^{2}$ Guangdong Provincial Key Laboratory of Quantum Engineering and Quantum Materials, Guangdong-Hong Kong Joint Laboratory of Quantum Matter, South China Normal University, Guangzhou 510006, China}
\affiliation{$^{3}$Key Laboratory of Atomic and Subatomic Structure and Quantum Control (MOE), Guangdong Basic Research Center of Excellence for Structure and Fundamental Interactions of Matter, Institute of Quantum Matter, South China Normal University, Guangzhou 510006, China}
\affiliation{$^{4}$Guangdong-Hong Kong Joint Laboratory of Quantum Matter, Guangdong Provincial Key Laboratory of Nuclear Science, Southern Nuclear Science Computing Center, South China Normal University, Guangzhou 510006, China}

%摘要
\begin{abstract}
The spherical geodesics of a neutral massive particle around Kerr-Newman black holes are investigated. 
The existence,  radii and radial stability of the equatorial and non-equatorial (particularly, the polar) spherical orbits are discussed for particles with different conserved energy. The radii of these orbits generally are solutions of a quintic polynomial equation with four dimensionless parameters, which are the rotation parameter $u$ and charge parameter $w$ of the black hole and the conserved angular momentum $\beta$ along the black hole rotation axis and conserved energy $\gamma$ of the particle. 
For the critical case where $\gamma=1$, we obtain the analytical expressions of the radii of the polar and equatorial orbits, and then derive the analytical radii for the general orbits. The radial stability of the orbits outside the event horizon is also discussed.  In the $(u, w, \beta)$ space, a no-orbit surface is found. When the parameters lies on this surface there is no orbit outside the event horizon, otherwise there is always one general spherical orbit outside the event horizon. 
For the cases with $0<\gamma<1$ and $\gamma>1$, we focus on the study of polar and equatorial orbits. For polar orbits with $0<\gamma<1$, a boundary surface in $(u, w, \gamma)$ space is identified which determines the existence of spherical polar orbits outside the event horizon. Numerical results of the radii and radial stability of the polar orbits are shown for examples with specific values of $\gamma$. For polar orbits with $\gamma>1$, it is found that there is always one unstable orbit outside the event horizon. 
For equatorial orbits with $0<\gamma<1$, in each rotating case (prograde case and retrograde case), a boundary surface in $(u, w, \gamma)$ space is also identified which divides the parameter space into two regions: one region with two orbits (one stable and the other unstable) and the other with no orbit outside the event horizon. Parameters on the boundary surface correspond to ISCOs. An analytical formula for the ISCOs is derived by choosing $(w,\gamma)$ as independent variables. For equatorial orbits with $\gamma>1$, it is found that there is always one unstable orbit outside the event horizon. Numerical results of the radii and radial stability of the equatorial orbits are also shown for examples with specific values of $\gamma$.
\end{abstract}
\maketitle

%正文
\section{Introduction}
Black holes are key objects studied in astrophysical and gravitational physics. The first observation of gravitational wave due to merger of black holes was reported by LIGO and Virgo in 2016 \cite{LIGOScientific:2016aoc}, which has opened a new era of astrophysics. Significant progress has also been made for observation of black holes via radio waves in recent years. The first image of the supermassive black hole M87* was released by Event Horizon Telescope (EHT) in 2019 \cite{EventHorizonTelescope:2019dse,EventHorizonTelescope:2019pgp,EventHorizonTelescope:2019ggy}. Subsequently, the image of the supermassive black hole Sagittarius A* in our galaxy center was also released\cite{EventHorizonTelescope:2022xnr}. These images provided us the spectacular appearance of the black holes with their accretion disks. 
In the theoretical study of these images, the null and timelike geodesics of the black holes play important roles. 

The study of the geodesics around rotating Kerr black holes is significantly simplified due to the discovery of Carter constant \cite{Carter:1968rr}.
In the equatorial plane of a Kerr black hole, two null geodesics with constant coordinate $r$ in Boyer-Lindquist coordinates exist outside the black hole horizon. The inner one  corresponds to a prograde orbit and the outer one corresponds to a retrograde orbit\cite{Bardeen:1972fi,Wilkins:1972rs,Teo:2003ltt}. 
For circular timelike geodesics in the equatorial plane, the spatial regions where they exist and their stability were discussed in \cite{Pugliese:2011xn}.  
A one-parameter class of solutions describing spherical null geodesics outside the black hole horizon was found in \cite{Teo:2003ltt}. An approximate formula for the radii of these spherical null geodesics was provided in \cite{Hod:2012ax}, and exact formulas for these spherical null geodesics have been discussed recently \cite{Tavlayan:2020cso}. A special spherical null geodesic, photon boomerang, is identified for nearly extremal Kerr black hole \cite{Page:2021rhx}. The general null geodesics of Kerr black holes can be expressed in terms of elliptic functions \cite{Gralla:2019ceu}.  
The existence of nonequatorial spherical timelike geodesics in extremal Kerr black holes was first discussed in \cite{Wilkins:1972rs}. Recent studies on the properties and radii of spherical timelike geodesics can be found in \cite{Teo:2020sey,Tavlayan:2021ylq}. 
Classification of radial timelike geodesic motion of the exterior nonextremal Kerr black hole was performed in \cite{Compere:2021bkk}.
General timelike geodesics of Kerr black holes can also be expressed in terms of elliptic functions \cite{Cieslik:2023qdc}. Kerr black holes admit both stable and unstable/plunging bound geodesics, separated in the parameter space by separatrix. The identification of the critical geodesics and the separatrix was studied in \cite{Levin:2008yp,Perez-Giz:2008ajn,Stein:2019buj}.  

Kerr-Newman black hole is the most general stationary, axisymmetric, asymptotically flat electrovacuum solution in four-dimensional Einstein-Maxwell theory.
The spherical timelike geodesics in Kerr-Newman black holes was first studied in \cite{Johnston:1974pn}. The properties of equatorial circular geodesics of neutral particles around Kerr-Newman black holes were studied in \cite{Dadhich1977,Pugliese:2013zma,Liu:2017fjx}. 
The study of timelike geodesics was also extended to nonequatorial cases for neutral \cite{Wang:2022ouq} and charged test particles \cite{Hackmann:2013pva}. 
Exact solutions of homoclinic orbits for the timelike geodesics of the particle on the general nonequatorial orbits in the Kerr-Newman black holes was presented in \cite{Li:2023bgn}.
The null trajectories and closed photon orbits in Kerr-Newman black holes were studied in \cite{Calvani:1981ya,Calvani:1980is,deVries:1999tiy}. 
A novel optical description of Kerr and Kerr-Newman black holes was provided in \cite{Galtsov:2019bty}.
The light bending in equatorial plane around Kerr-Newman black holes was studied in \cite{Hsiao:2019ohy}, which was also extended to the extremal Kerr-Newman case\cite{Chen:2024oyv}.  
 Recently, interesting topological arguments have been proposed to prove the existence of planar bound null geodesics around stationary, axisymmetric spacetimes \cite{Cunha:2020azh,Guo:2020qwk,Wei:2020rbh,Ghosh:2021txu}.

 Recently, the spherical timelike geodesics of a neutral test particle around Kerr-Newman black holes have been discussed and the radius of innermost stable spherical orbit is plotted as a function of the Carter constant \cite{Alam:2024mmw}. However, unlike the Kerr black hole case, the radii of spherical timelike orbits around a Kerr-Newman black hole are still lack of systematical study. In a previous study, we discussed the spherical null geodesics around Kerr-Newman black holes, and provided several exact formulas for the radii of these geodesics \cite{Chen:2022ewe}.
In this work, we consider the spherical timelike geodesics of a neutral test particle around a Kerr-Newman black hole, focus on the number of orbits outside the event horizon and the analytical formulas for the radii of orbits as functions of parameters of the black hole and neutral particle.

The paper is organized as follows. In Section II, we review the timelike geodesics around a Kerr-Newman black hole and derive a quintic polynomial equation satisfied by the radii of the spherical timelike orbits. In Section III, we consider the existence and radii of the  spherical orbits outside the black hole event horizon for the critical case where the conserved energy parameter $\gamma$ of the particle equals to 1.  We also consider the radii of spherical orbits for different limiting cases of the Kerr-Newman black hole and check
 our results with previous ones in literature. In Section IV, we consider the radii of the timelike spherical orbits for the non-critical cases with $\gamma>1$  and $0<\gamma<1$. The last section is devoted to the conclusion.

\section{Timelike geodesics of Kerr-Newman black holes}\label{sec.2}
The metric of a Kerr-Newman black hole with mass \( M \), angular momentum \( J = Ma \), and electric charge \( Q \) in Boyer-Lindquist coordinates is expressed as follows (we adopt geometric units where $G = c = 1$)
\begin{equation}
	\begin{split}
		\mathrm{d}s^2=&-\frac{\Delta}{\Sigma}(\mathrm{d}t-a\sin^2{\theta}\mathrm{d}\phi)^2+\frac{\Sigma}{\Delta}\mathrm{d}r^2+\Sigma\mathrm{d}\theta^2\\
		&+\frac{\sin^2{\theta}}{\Sigma}[(r^2+a^2)\mathrm{d}\phi-a\mathrm{d}t]^2\ ,
	\end{split}
\end{equation}
where 
\begin{equation}
	\Sigma=r^2+a^2\cos^2{\theta}\ ,\quad \Delta=r^2-2Mr+a^2+Q^2\ .
\end{equation}
The outer and inner horizons are the roots of $\Delta=0$, which are
\begin{equation}
	r_{\pm}=M\pm\sqrt{M^2-a^2-Q^2}\ .
\end{equation} 

Now we consider a neutral massive particle moving in the Kerr-Newman black hole. For simplicity, we assume the massive particle has unit mass. The Lagrangian of the particle is 
\begin{equation}
	\mathcal{L}=\frac{1}{2}g_{\mu\nu}u^\mu u^\nu\ ,
\end{equation} 
where \( u^\mu = \frac{\mathrm{d}x^\mu}{\mathrm{d}\tau} \) is the four-velocity of the particle, and \( \tau \) is the proper time. For a particle in Kerr-Newman spacetime, there are two Killing vectors, \( \xi_{t} = \frac{\partial}{\partial t} \) and \( \xi_{\phi} = \frac{\partial}{\partial \phi} \), which yield two conserved quantities along the geodesics. The first is the energy \( E = -p_t = -\frac{\partial\mathcal{L}}{\partial \dot{t}} \), representing the energy per unit mass measured by a static observer at infinity. The second is the orbital angular momentum per unit mass along the symmetric axis of the black hole, \( L = p_\phi = \frac{\partial\mathcal{L}}{\partial \dot{\phi}} \). In fact, there exists another important constant of motion, \( C \), known as the Carter constant, which governs the particle's motion in the \( \theta \)-direction\cite{Carter:1968rr,Teo:2003ltt}, 
\begin{equation}
	C=\Sigma^2(u^\theta)^2-a^2E^2\cos^2\theta+L^2\cot^2{\theta}+a^2\cos^2{\theta}\ .
\end{equation}
$C=0$ means the equatorial motion of particle\cite{Hod:2012ax,Tavlayan:2020cso}.

Using the three conserved quantities mentioned above and applying the four-velocity normalization \( u^\mu u_\mu = -1 \), we can derive following four first-order equations of geodesic motion for a massive particle in Kerr-Newman spacetime,
\begin{equation}
    \begin{split}
        \Sigma\frac{\mathrm{d}r}{\mathrm{d}\tau}&=\pm \sqrt{R(r)},\\
        \Sigma\frac{\mathrm{d}\theta}{\mathrm{d}\tau}&=\pm \sqrt{\Theta(\theta)},\\
        \Sigma\frac{\mathrm{d}\phi}{\mathrm{d}\tau}&=\frac{a}{\Delta}[(r^2+a^2)E-aL]+(\frac{L}{\sin^2{\theta}}-aE),\\
        \Sigma\frac{\mathrm{d}t}{\mathrm{d}\tau}&=\frac{r^2+a^2}{\Delta}[(r^2+a^2)E-aL]+a(L-aE\sin^2{\theta})\ ,
    \end{split}
\end{equation}
where
\bea \nn
R(r)&=&[(r^2+a^2)E-aL]^2-\Delta[C+(L-aE)^2+r^2],\label{R(r)}\\	
\Theta(\theta)&=&C+a^2\cos^2{\theta}(E^2-1)-L^2\cot^2{\theta}.    
\eea
The geodesic motions of the massive particle can be divided into three cases with respect to the value of the conserved energy \( E \),
\begin{itemize}
    \item \( 0 < E < 1 \): The particle does not have enough energy to escape to infinity and remains in a bounded motion;
    \item \( E = 1 \): This is a critical case and the particle has just enough energy to escape to infinity;
    \item \( E > 1 \): The particle has enough energy to escape to infinity.
\end{itemize}

In this work, we focus on the timelike spherical orbits with a constant radius \( r \) outside the event horizon. These orbits are determined by the following two equations\cite{Teo:2003ltt,Hod:2012ax,Tavlayan:2020cso,Bardeen:1972fi},
\begin{equation}\label{r=0}
	R(r)=0\ ,\quad \frac{\mathrm{d}R(r)}{\mathrm{d}r}=0\ .
\end{equation} 
Plugging equation \eqref{R(r)} into the above equations, we can solve the Carter constant from each equation and obtain
\begin{align}
	\begin{split}
		C=&\frac{1}{M-r}(-a^2 E^2 M-a^2 E^2 r+a^2 r+2 a
		L E M\\
		&-L^2 M+L^2 r-2E^2 r^3-3 M r^2+Q^2 r+2 r^3)\ ,
	\end{split}\label{c1}\\
	\begin{split}
		C=&\frac{1}{-a^2+2 M r-Q^2-r^2}(-2 a^2 E^2 M r+a^2 E^2 Q^2\\
		&-a^2E^2 r^2+a^2 r^2+4 a L E M r-2 aL E Q^2-2 L^2 M r\\
		&+L^2Q^2+L^2 r^2-E^2 r^4-2 M r^3+Q^2r^2+r^4)\ .\label{c2}
	\end{split}
\end{align}
To simplify the analysis of the above equations, we introduce the following dimensionless parameters
\begin{equation}\label{diml}
	x\equiv\frac{r}{M}, u\equiv\frac{a}{M}, w\equiv\frac{Q}{M}, \beta\equiv\frac{L}{M}, c\equiv\frac{C}{M^2}, E=\gamma\ .
\end{equation}
These parameters satisfy the following  physical constraints (since \( Q \) appears only in square, we just consider the case with positive charge),
\begin{equation}\label{phyvalue}
	\begin{split}
		x&\geq0\ , \quad 0\leq u\leq 1\ ,\quad 0\leq w \leq 1\ ,\\
		0&\leq u^2+w^2\leq 1\ ,\quad \beta \in \mathbb{R}\ .
	\end{split}
\end{equation}

The right-hand sides of equations \eqref{c1} and \eqref{c2} must be equal, then we obtain the following quintic equation,
\begin{equation}\label{P5}
	\begin{split}
	P_5(x)=&\left(\gamma ^2-1\right) x^5+\left(4-3\gamma ^2\right) x^4\\
	&+\left(2\gamma ^2 u^2-2 u^2+2 \gamma ^2 w^2-2 w^2-4\right)x^3\\
	&+\left(-2 \gamma ^2 u^2+4 u^2+2 \beta  \gamma  u+4w^2\right)x^2\\
	&+\left(\gamma ^2 u^4-u^4-\beta ^2 u^2+2 \gamma^2 u^2 w^2\right.\\
	&\left.-2 u^2 w^2-2 \beta  \gamma  uw^2-w^4\right)x\\
	&+\gamma ^2 u^4-2 \beta  \gamma  u^3+\beta ^2 u^2=0 .
	\end{split}
\end{equation}
The positive real roots of this equation yield the radii of the spherical orbits. Furthermore, by considering the massless limit (\( \gamma \rightarrow \infty \)), we can arrive at the same equation for the spherical photon orbit\cite{Chen:2022ewe}.

\section{Radii of spherical timelike orbits with $\gamma=1$}
In this section we consider the case with $\gamma=1$, which is the critical case. For this case, equation \eqref{P5} reduces to a quartic equation,
\begin{equation}\label{P4}
	\begin{split}
		&x^4-4 x^3+\left(2u^2+2 \beta  u+4 w^2\right)x^2\\
		&-\left(\beta ^2 u^2+2 \beta uw^2+w^4\right)x \\
		&+u^4-2 \beta  u^3+\beta ^2 u^2=0\ .
	\end{split}
\end{equation}

\subsection{Some limiting cases}
By setting the rotation parameter \( u = 0 \) and the charge parameter \( w = 0 \), the Kerr-Newman black hole reduces to a Schwarzschild black hole. Ignoring the zero roots, the quartic equation \eqref{P4} reduces to
\begin{equation}
	x-4=0\ .
\end{equation}
It is straightforward to see that the above equation has only one solution \( x = 4 \), indicating that there is a spherical orbit with a radius of \( x = 4 \) (\( r = 4M \)) around a Schwarzschild black hole. The horizon of the Schwarzschild black hole is located at \( r = 2M \), meaning that the spherical orbit lies outside the horizon. We define 
\begin{equation}\label{rq}
	\tilde{R}(x)=\frac{R(x)}{M^4}\ ,\quad \tilde{R}^{(2)}_i=\frac{\mathrm{d}^2\tilde{R}(x)}{\mathrm{d}x^2}\mid_{x=x_i}\ .
\end{equation}
For \( x_i = 4 \), we find that \( \tilde{R}^{(2)}_i = 16 > 0 \), which indicates that the orbit is unstable under radial perturbation. This unstable orbit is referred as the inner bound circular orbit (IBCO)\cite{Barack:2019agd}.

\begin{figure}[H] % H为当前位置，!htb为忽略美学标准，htbp为浮动图形
    \centering % 图片居中
    \includegraphics[width=0.4\textwidth]{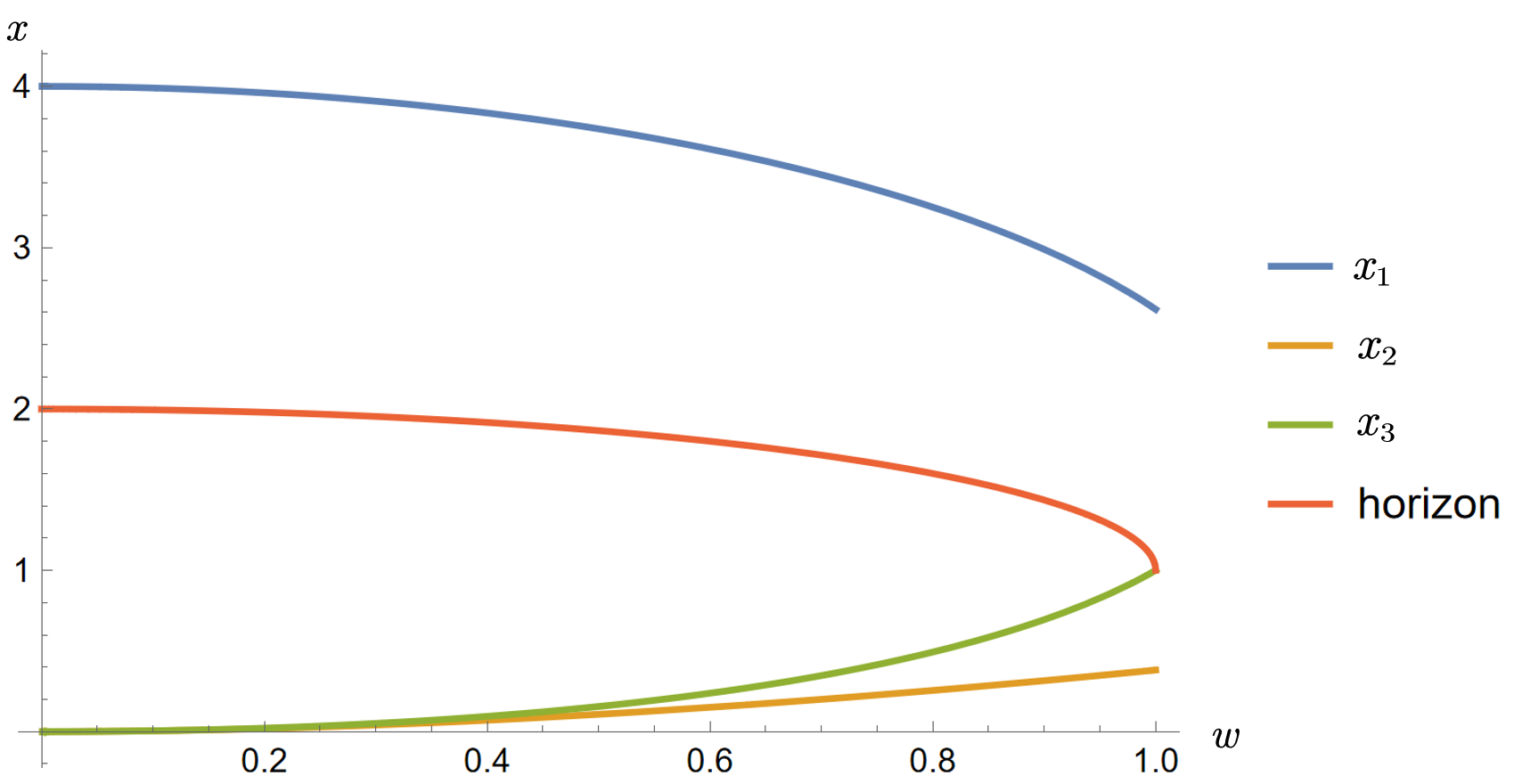} % 插入图片，[]中设置图片大小，{}中是图片文件名
    \caption{Radii of the spherical timelike orbits and the horizon of RNBH as functions of $w$.} % 最终文档中希望显示的图片标题
    \label{rnsol} % 用于文内引用的标签
\end{figure}

\begin{figure}[H] % H为当前位置，!htb为忽略美学标准，htbp为浮动图形
    \centering % 图片居中
    \includegraphics[width=0.4\textwidth]{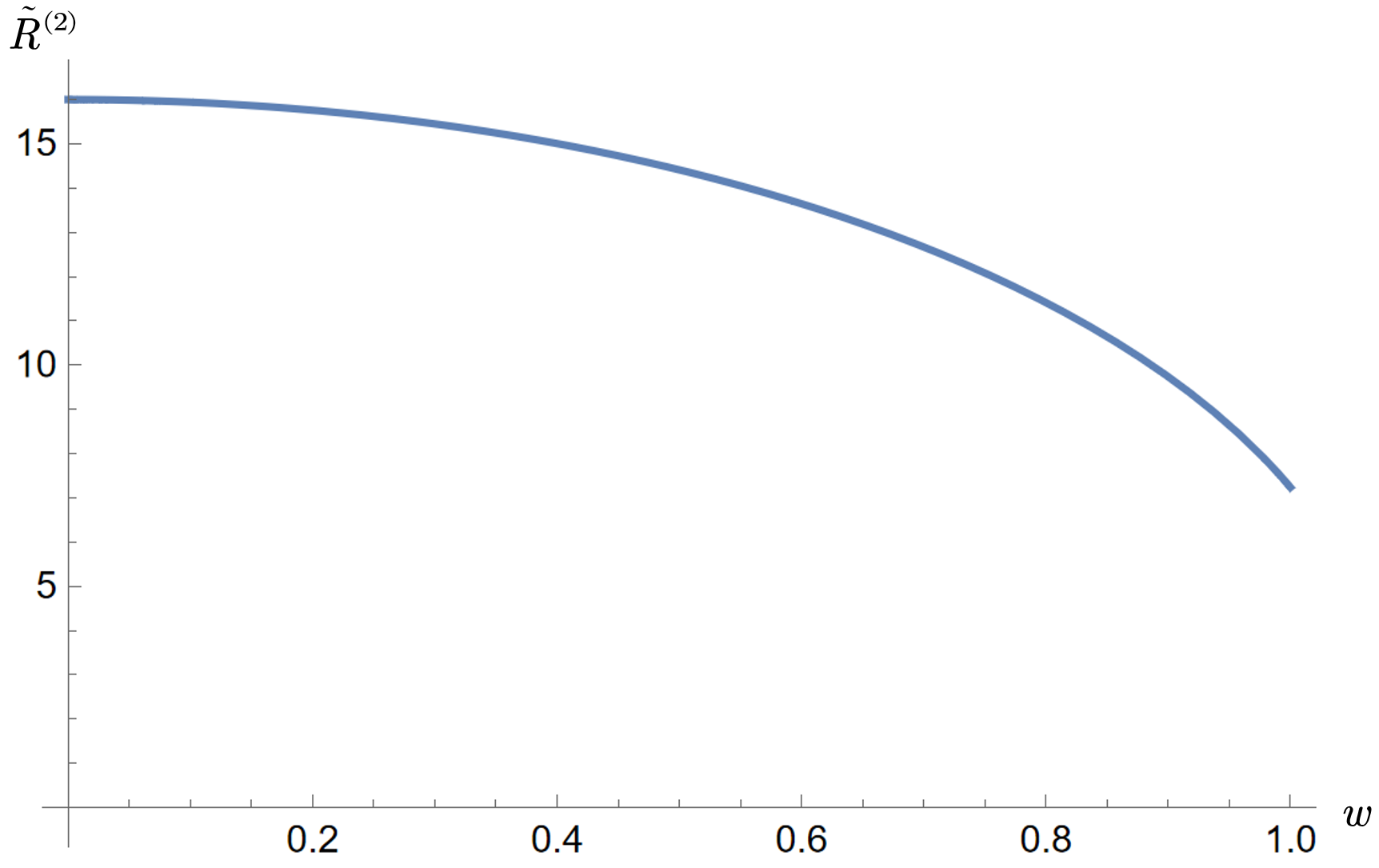} % 插入图片，[]中设置图片大小，{}中是图片文件名
    \caption{$\tilde{R}^{(2)}_i$ is always positive.} % 最终文档中希望显示的图片标题
    \label{rnrq} % 用于文内引用的标签
\end{figure}

When we set \( u = 0 \) while allowing \( w \neq 0 \), the Kerr-Newman black hole reduces to a Reissner-Nordström black hole (RNBH). The quartic equation \eqref{P4} then reduces to (ignoring the zero root)
\begin{equation}
	x^3-4x^2+4w^2x-w^4=0\ ,
\end{equation}
and its three real roots are
\begin{equation}
    x_{k}=\frac{4}{3}+2\sqrt{-\frac{p}{3}}\cos\left(\frac{\theta+2(k-1)\pi}{3}\right)\ ,\ k=1,2,3\ ,
\end{equation}
where 
\begin{equation}
    \begin{split}
        \theta&=\arccos\left(\frac{q}{2}\sqrt{-\frac{27}{p^3}}\right)\ ,\quad p=\frac{1}{3}(12w^2-16)\ ,\\
        q&=\frac{1}{27}(27w^4-144w^2+128)\ .
    \end{split}
\end{equation}
The radii of the spherical orbits are functions of the charge parameter \( w \). Given the horizon radius of the RNBH, \( x_h = 1 + \sqrt{1 - w^2} \), we plot the three radii and the horizon radius as functions of \( w \) in Fig.\ref{rnsol}.
 % 结束环境
Thus, there is one spherical orbit with radius \( x_1 \) located outside the horizon. According to Eq.\eqref{rq} and taking \( x_i = x_1 \), we find that \( \tilde{R}^{(2)}_i \) is a function of \( w \) and is always positive, which is illustrated in Fig.\ref{rnrq}. Therefore, the spherical orbit with radius \( x_1 \) is unstable under radial perturbation.
 % 结束环境

\begin{figure}[H] % H为当前位置，!htb为忽略美学标准，htbp为浮动图形
    \centering % 图片居中
    \includegraphics[width=0.4\textwidth]{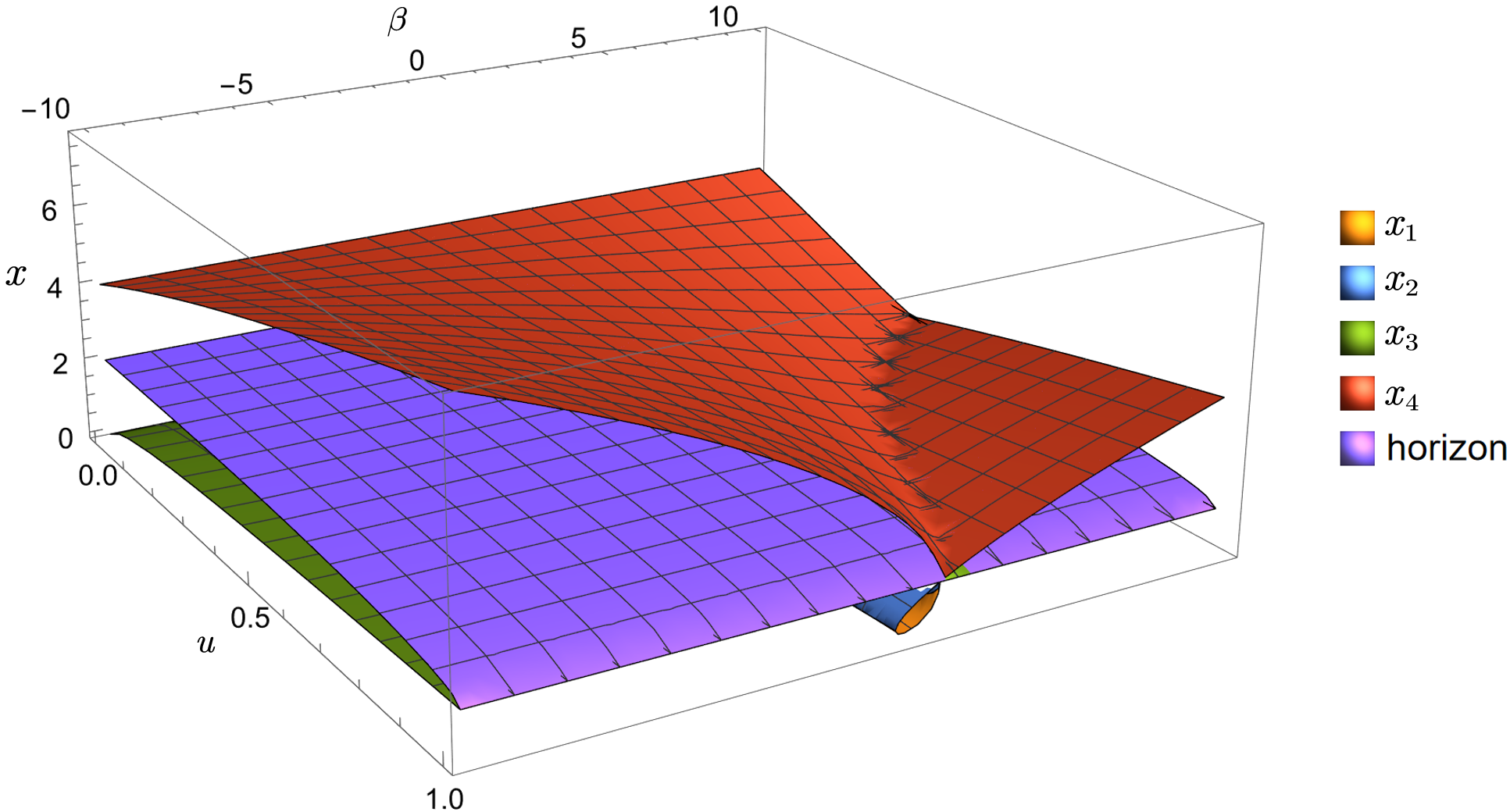} % 插入图片，[]中设置图片大小，{}中是图片文件名
    \caption{Radii of four spherical orbits and the horizon of Kerr black hole as functions of $u$ and $\beta$.} % 最终文档中希望显示的图片标题
    \label{kerrgesol} % 用于文内引用的标签
\end{figure}

When we set $u\neq 0$ while allowing $w=0$, the Kerr-Newman black hole becomes a Kerr black hole. The equation \eqref{P4} reduces to 
\begin{equation}
    \begin{split}
        &x^4-4x^3+(2u^2+2u\beta)x^2-u^2\beta^2 x\\
        &+u^2\beta^2-2u^3\beta+u^4=0\ .
    \end{split}
\end{equation}
The analytical expression of the four roots of the above quartic equation can be obtained from the general formula in the \hyperref[appendix]{Appendix}. Here we plot the four roots (when they are real) and the black hole outer horizon $x_h=1+\sqrt{1-u^2}$ as functions of $u$ and $\beta$ in Fig.\ref{kerrgesol}.
 % 结束环境
We can see that the four roots are not always real for various values of the parameters $u$ and $\beta$ and there is only one orbit with radius $x_4$ outside the event horizon. For each set of values of $u$ and $|\beta|$, there is a prograde orbit with $\beta>0$ and a retrograde orbit with $\beta<0$, which is consistent with the results in \cite{Tavlayan:2021ylq}. 

It is  noticed that there is a curve in ($u,\beta$)-plane for which the prograde orbit coincides with the event horizon. We find that the equation of this curve in ($u,\beta$)-plane is 
 \bea
 u=\frac{4\beta}{\beta^2+4},~\beta\geq2.
 \eea

\subsection{Spherical orbits in Kerr-Newman black hole}\label{secb}
In this subsection we consider the radii of spherical timelike orbits in Kerr-Newman black hole for a neutral massive particle with $\gamma=1$.
\subsubsection{Polar orbits}
\begin{figure}[H] % H为当前位置，!htb为忽略美学标准，htbp为浮动图形
    \centering % 图片居中
    \includegraphics[width=0.4\textwidth]{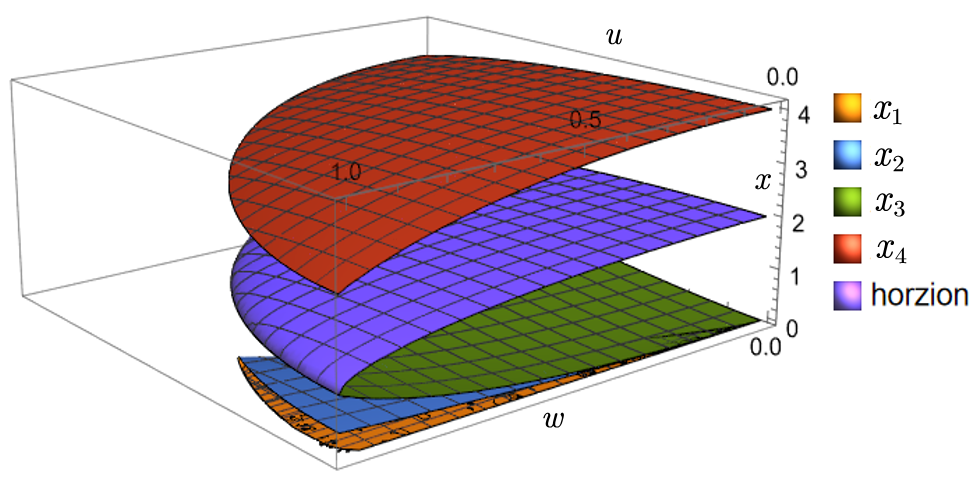} % 插入图片，[]中设置图片大小，{}中是图片文件名
    \caption{Radii of four polar orbits and radius of the event horizon are plotted as functions of $u$ and $w$.} % 最终文档中希望显示的图片标题
    \label{knpsol} % 用于文内引用的标签
\end{figure}

\begin{figure}[H] % H为当前位置，!htb为忽略美学标准，htbp为浮动图形
    \centering % 图片居中
    \includegraphics[width=0.4\textwidth]{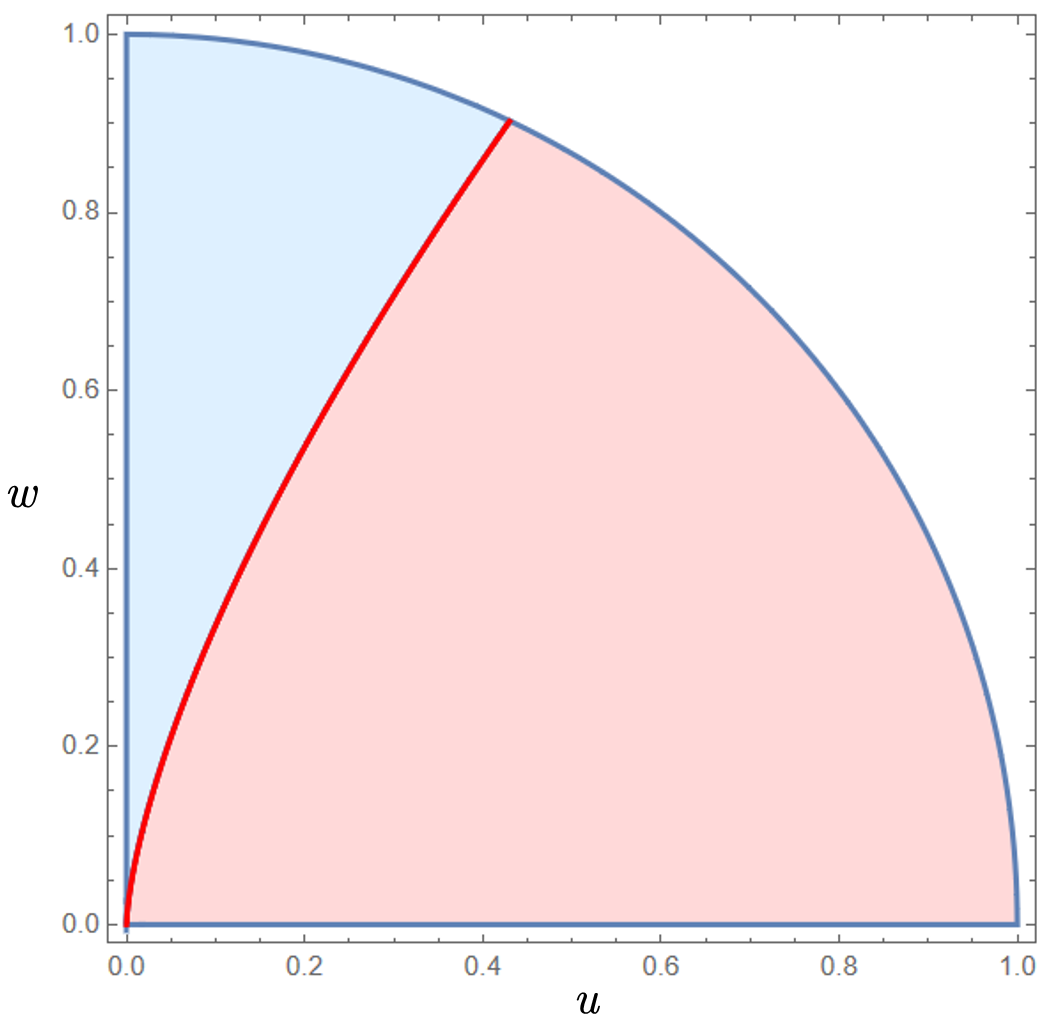} % 插入图片，[]中设置图片大小，{}中是图片文件名
    \caption{Two regions (light blue and light red) and their separating curve (red) in $(u,w)$-plane.} % 最终文档中希望显示的图片标题
    \label{knpc} % 用于文内引用的标签
\end{figure}

\begin{figure}[H] % H为当前位置，!htb为忽略美学标准，htbp为浮动图形
    \centering % 图片居中
    \includegraphics[width=0.4\textwidth]{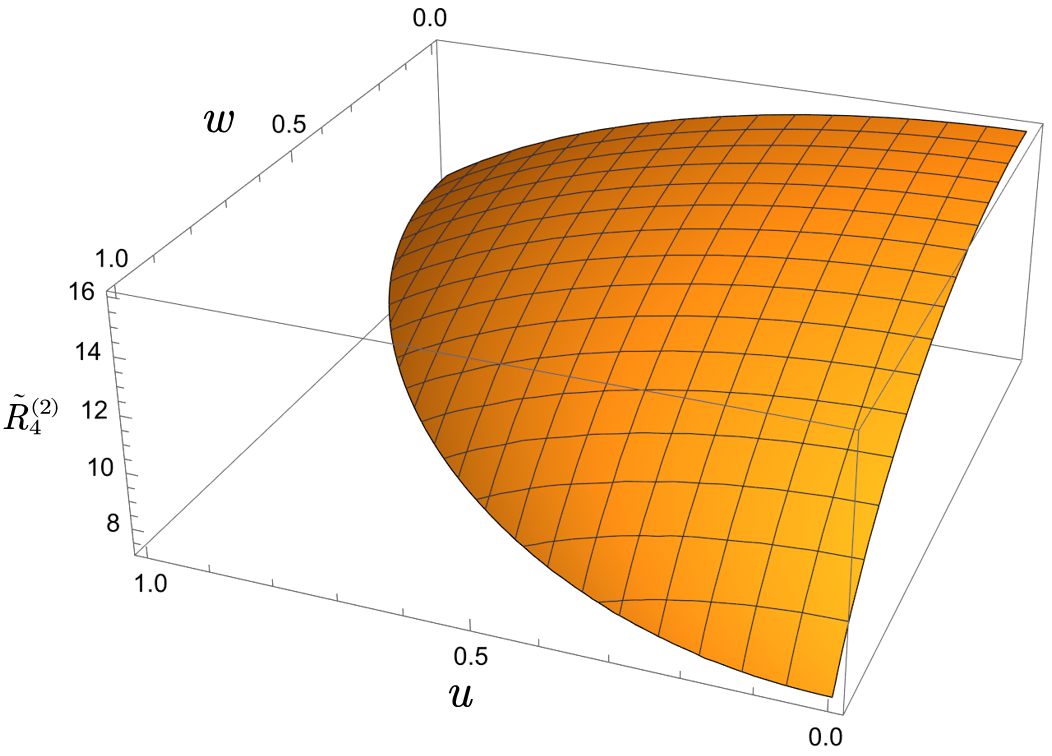} % 插入图片，[]中设置图片大小，{}中是图片文件名
    \caption{$\tilde{R}^{(2)}_4$ is always positive as a function of $u$ and $w$.} % 最终文档中希望显示的图片标题
    \label{knprq} % 用于文内引用的标签
\end{figure}

For the spherical polar orbits, the angular momentum parameter of the particle vanishes, i.e. $\beta=0$ ($L=0$). Then, the equation \eqref{P4} reduces to the following quartic equation, 
\begin{equation}\label{pp4}
	x^4-4x^3+(2u^2+4w^2)x^2-w^4x+u^4=0 . 
\end{equation}
The analytical expressions of its four roots can be obtained from the general formula in the \hyperref[appendix]{Appendix}.
Given the expression of the event horizon of the Kerr-Newman black hole \( x_h = 1 + \sqrt{1 - u^2 - w^2} \), we plot the four roots and the event horizon as functions of \( u \) and \( w \) in Fig.\ref{knpsol}.
 % 结束环境
We find that the number of orbits varies with different values of \( u \) and \( w \). There are either two or four real roots for equation \eqref{pp4} in two different regions of the \( (u, w) \)-plane. The curve separating these two regions is determined by the vanishing of the discriminant of Eq.\eqref{pp4},
\bea
        &&4096 u^{10}+2048 u^8 w^4+6144 u^8 w^2-6912 u^8+256 u^6w^8\nn\\
		&&+3072 u^6 w^6-3840 u^6 w^4+384 u^4 w^{10}-1056 u^4 w^8\nn\\
		&&+512 u^4 w^6-240 u^2 w^{12}+256 u^2 w^{10}-27
        w^{16}+32 w^{14}\nn\\
        &&=0 .
\eea
We plot the two regions and the curve in the \( (u, w) \)-plane in Fig.\ref{knpc}. When  \( u \) and \( w \) belong to the light blue region,  there are four real roots to Eq.\eqref{pp4}. When  \( u \) and \( w \) belong to the light red region, there are two real roots to Eq.\eqref{pp4}. The red curve separates these two regions.
 % 结束环境
From Fig.\ref{knpsol}, it is also noticed that only one spherical polar orbit with radius \( x_4 \) is outside the event horizon. 
 According to Eq.\eqref{rq} and taking \( x_i = x_4 \), we plot \( \tilde{R}^{(2)}_4 \) as a function of \( u \) and \( w \) in Fig.\ref{knprq}. It is easy to see that it is always positive, which indicates that the spherical orbit is unstable under radial perturbation.
 % 结束环境

\subsubsection{Equatorial orbits}
Spherical orbits constrained on the black hole equatorial plane are circular orbits. In this case, the Carter constant \( C = 0 \), i.e.,  Eqs.\eqref{c1} and \eqref{c2} vanish. 
Now, there are three parameters in Eqs.\eqref{c1} and \eqref{c2}, charge parameter $w$ and rotation parameter $u$ of the black hole and angular momentum parameter $\beta$ of the massive particle. 
Usually, we can choose the black hole parameters as free parameters and solve Eqs.\eqref{c1} and \eqref{c2} for the orbit radius $r$ and particle parameter $\beta$. However, after canceling $\beta$, the equation to determine $x$ is non-polynomial and complex. 

Here, we choose the rotation parameter $u$ and angular momentum parameter $\beta$ as free parameters, and solve Eqs.\eqref{c1} and \eqref{c2} for the orbit radius $r$ and parameter $\omega$.  Given the vanishing of Eqs.\eqref{c1} and \eqref{c2}, we obtain  
\begin{align}
	\begin{split}
		w^2=&\frac{1}{(\beta-\gamma u )^2+x^2}(\gamma ^2u^2 x^2-u^2 x^2+2 \gamma ^2 u^2 x-x^4\\
		&-4 \beta \gamma  u x+\gamma ^2 x^4+2 x^3-\beta ^2 x^2+2 \beta ^2 x)\ ,
	\end{split}\label{ws1}\\
	\begin{split}
		w^2=&\frac{1}{x}(\beta^2+\gamma ^2 u^2+\gamma ^2 u^2 x-u^2 x-2\beta  \gamma u\\
		&+2 \gamma ^2 x^3-2 x^3+3x^2-\beta ^2 x)\ ,
	\end{split}\label{ws2}
\end{align}
where the parameter $\gamma$ is general and temporarily not fixed to 1.
The righthand sides of Eqs.\eqref{ws1} and \eqref{ws2} must be equal, leading to a quintic equation
\begin{equation}\label{peq}
	\begin{split}
	&\left(1-\gamma ^2\right) x^5-x^4+ \left(-2\beta ^2 \gamma ^2+2 \beta ^2-2 \gamma ^4u^2+2 \gamma ^2 u^2\right.\\
	&\left.+4 \beta  \gamma ^3 u-4\beta  \gamma  u\right)x^3+ \left(-2 \beta^2-2 \gamma ^2 u^2+4 \beta  \gamma u\right)x^2\\
	&+ \left(\beta ^4-\gamma ^4u^4+\gamma ^2 u^4+2 \beta  \gamma ^3 u^3-2\beta  \gamma  u^3+\beta ^2 u^2\right.\\
	&\left.-2 \beta ^3\gamma  u\right)x
	-\beta ^4-\gamma ^4 u^4+4 \beta  \gamma ^3u^3-6 \beta ^2 \gamma ^2 u^2\\
	&+4 \beta ^3 \gamma u=0.
	\end{split}
\end{equation}
Consider $\gamma=1$, the above equations reduces to 
\begin{align}
	\begin{split}
		&-x^4+(-2u^2+4u\beta-2\beta^2)x^2+(u^2\beta^2-2u\beta^3\\
		&+\beta^4)x
		-u^4+4u^3\beta-6u^2\beta^2+4u\beta^3-\beta^4=0\ ,
	\end{split}\label{peqc}\\
	\begin{split}
		w^2=\frac{\beta ^2+u^2-2 \beta  u+3 x^2-\beta ^2x}{x}\ .
	\end{split}	\label{wsc}
\end{align}
The solutions of Eq.\eqref{peqc} are expressed in terms of \( u \) and \( \beta \), their analytic forms can be found in \hyperref[appendix]{Appendix}. It should be noted that for different signs of \( \beta \), we need to classify the orbits into prograde orbits ($\beta>0$) and retrograde orbits ($\beta<0$).

\begin{figure}[H] % H为当前位置，!htb为忽略美学标准，htbp为浮动图形
    \centering % 图片居中
    \includegraphics[width=0.4\textwidth]{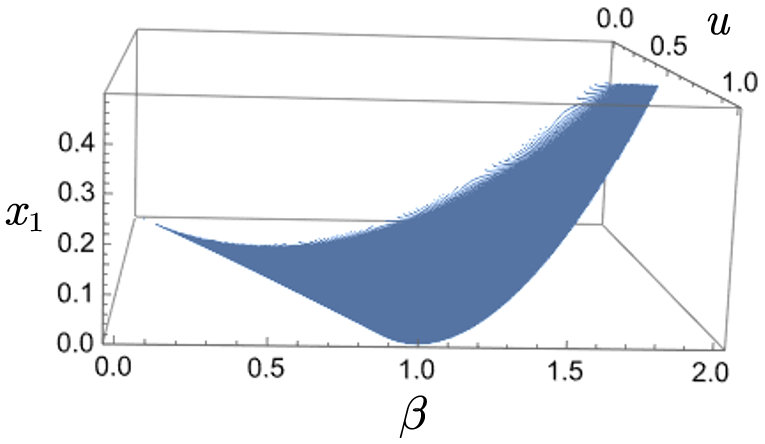} % 插入图片，[]中设置图片大小，{}中是图片文件名
    \caption{Orbit radius $x_1$ as a function of $u$ and $\beta$.} % 最终文档中希望显示的图片标题
    \label{kneqsol1} % 用于文内引用的标签
\end{figure}

\begin{figure}[H] % H为当前位置，!htb为忽略美学标准，htbp为浮动图形
    \centering % 图片居中
    \includegraphics[width=0.4\textwidth]{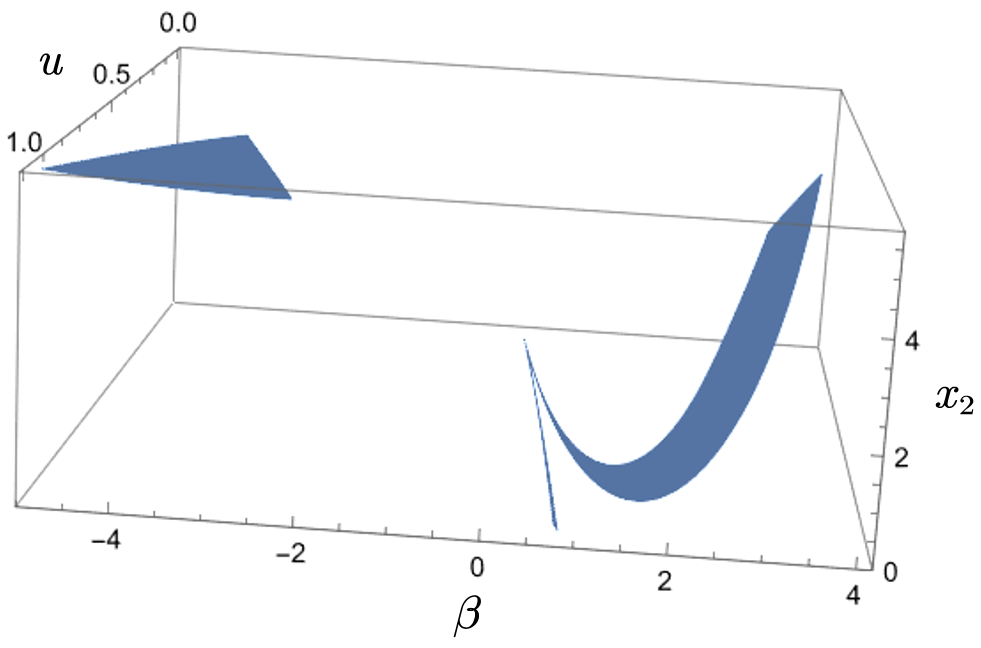} % 插入图片，[]中设置图片大小，{}中是图片文件名
	\caption{ Orbit radius $x_2$ as a function of $u$ and $\beta$.} % 最终文档中希望显示的图片标题
    \label{kneqsol2} % 用于文内引用的标签
\end{figure}

\begin{figure}[H] % H为当前位置，!htb为忽略美学标准，htbp为浮动图形
    \centering % 图片居中
    \includegraphics[width=0.4\textwidth]{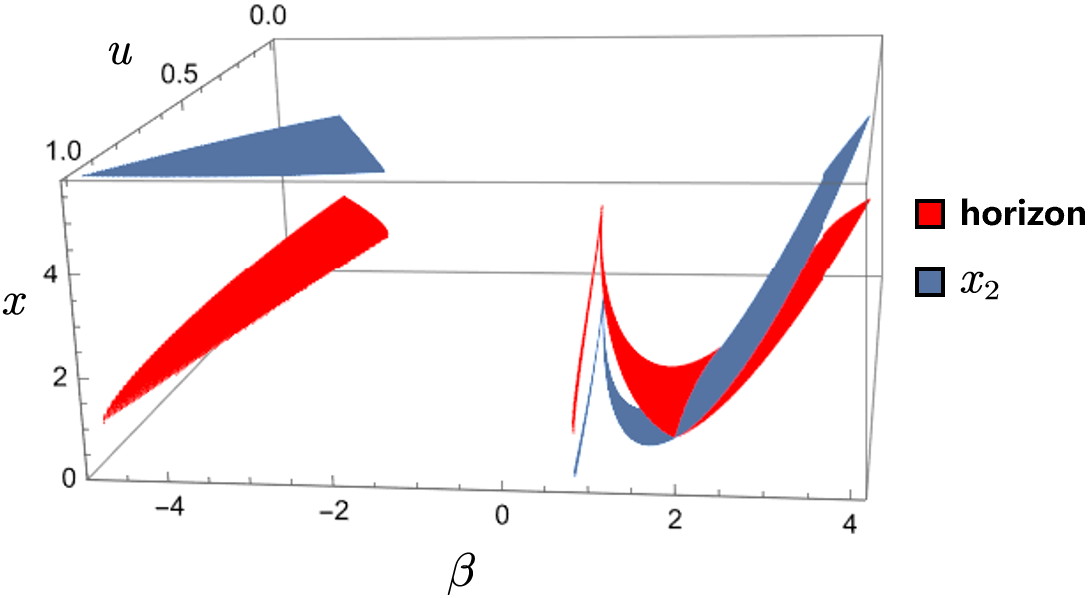} % 插入图片，[]中设置图片大小，{}中是图片文件名
    \caption{Orbit with radius $x_2$ is outside the event horizon for two parameter regions in \( (u, \beta) \)-plane.} % 最终文档中希望显示的图片标题
    \label{kneqsol3} % 用于文内引用的标签
\end{figure}

\begin{figure}[H] % H为当前位置，!htb为忽略美学标准，htbp为浮动图形
    \centering % 图片居中
    \includegraphics[width=0.4\textwidth]{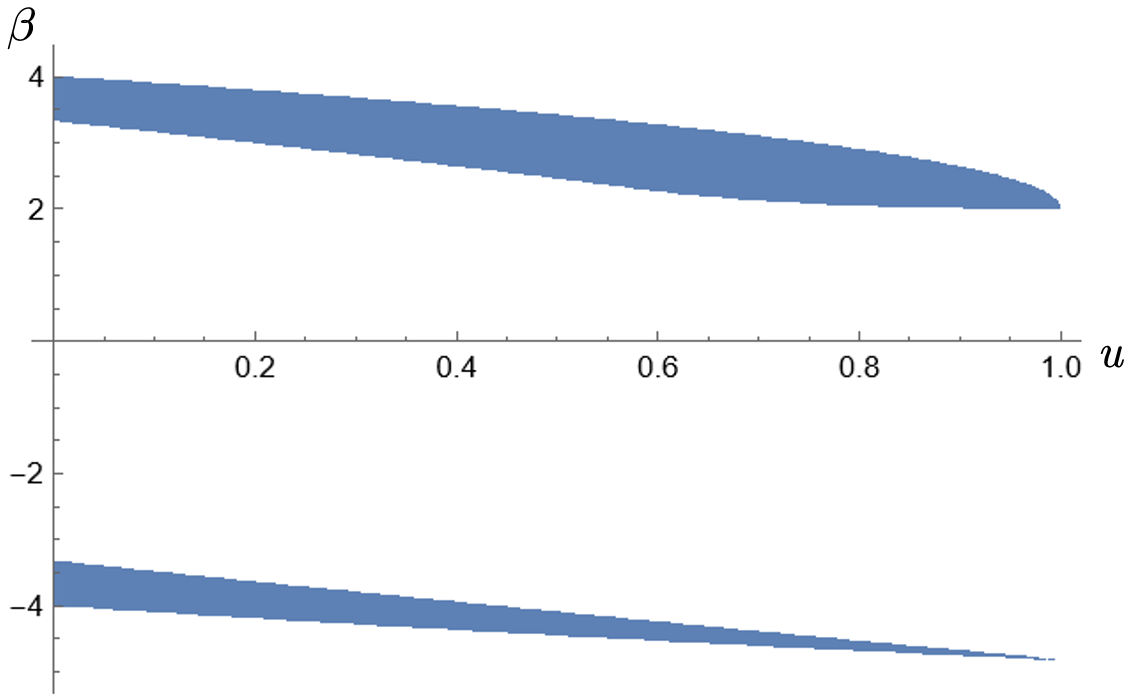} % 插入图片，[]中设置图片大小，{}中是图片文件名
    \caption{Two parameter regions where orbit radius $x_2$ is larger than the event horizon radius $x_h$.} % 最终文档中希望显示的图片标题
    \label{kneqpara} % 用于文内引用的标签
\end{figure}

Using numerical calculation, it is found that \( x_3 \) and \( x_4 \) are not real numbers within the physically allowed ranges \( 0 \leq u^2 \leq 1, \ 0 \leq w^2 \leq 1, \ 0 \leq u^2 + w^2 \leq 1 \). Therefore, we just focus on \( x_1 \) and \( x_2 \). The two roots are plotted as functions of \( u \) and \( \beta \), and shown in Fig.\ref{kneqsol1} and Fig.\ref{kneqsol2}, respectively. The orbit with radius $x_1$ is always prograde, and orbit with radius $x_2$ may be prograde or retrograde.
 % 结束环境
 % 结束环境
The event horizon radius \( x_h = 1 + \sqrt{1 - u^2 - w^2} \) satisfies \( 1 \leq x_h \leq 2 \). Therefore, the orbit with radius \( x_1 \) lies inside the event horizon.
In Fig.\ref{kneqsol3}, we can see that the orbit with radius \( x_2 \) is outside the event horizon for two specific parameter regions in the \( (u, \beta) \)-plane. 
The two parameter regions in the \( (u, \beta) \)-plane where \( x_2 >x_h\) is numerically plotted in Fig.\ref{kneqpara}. These two regions correspond to prograde ($\beta>0$) and retrograde ($\beta<0$) orbits, respectively.

 % 结束环境

\begin{figure}[H] % H为当前位置，!htb为忽略美学标准，htbp为浮动图形
    \centering % 图片居中
    \includegraphics[width=0.4\textwidth]{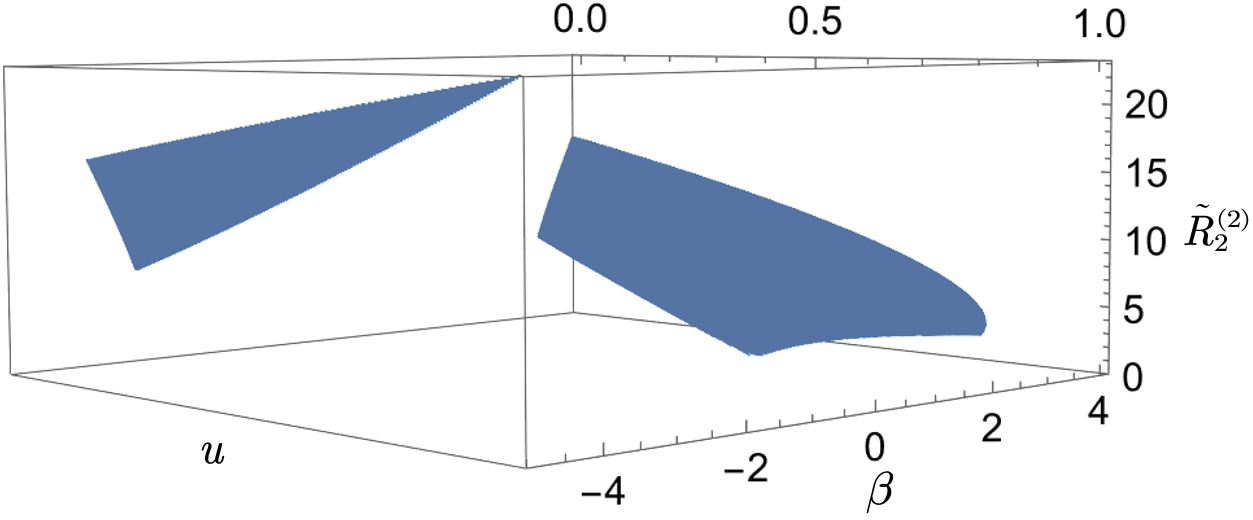} % 插入图片，[]中设置图片大小，{}中是图片文件名
    \caption{$\tilde{R}^{(2)}_2>0$ in this region.} % 最终文档中希望显示的图片标题
    \label{kneqr} % 用于文内引用的标签
\end{figure}

 % 结束环境
With Eq.\eqref{rq}, we can analyze the stability of the orbit with radius \( x_2 \) in these regions by calculating its \( \tilde{R}^{(2)}_2 \), which is a function of \( u \) and \( \beta \). We find that \( \tilde{R}^{(2)}_2 > 0 \) in these regions, which can be seen in Fig.\ref{kneqr}. This means that the prograde and retrograde orbits are both unstable.
 % 结束环境

\subsubsection{General cases}
In this section, we consider the most general case for spherical timelike orbits around Kerr-Newman black holes. The equation satisfied by the orbit radius is given by equation \eqref{P4}, and its solutions are functions of \( u \), \( w \), and \( \beta \), their analytic forms can also be obtained with formula in the \hyperref[appendix]{Appendix}.

In Fig.\ref{kngsol}, we fix the rotation parameter $u=0.5$ and the charge parameter $w=0.3$, and plot the four orbit radii and the event horizon radius as functions of the angular momentum parameter $\beta$.
We can see that orbits with radii $x_1,x_2$ only exist in a small interval around $\beta=0.5$ while orbits with radii $x_3,x_4$ always exist. 
It can also be noticed that only one orbit with radius $x_4$ is outside the event horizon, which can be checked to be unstable under radial perturbation. 

\begin{figure}[H] % H为当前位置，!htb为忽略美学标准，htbp为浮动图形
    \centering % 图片居中
    \includegraphics[width=0.38\textwidth]{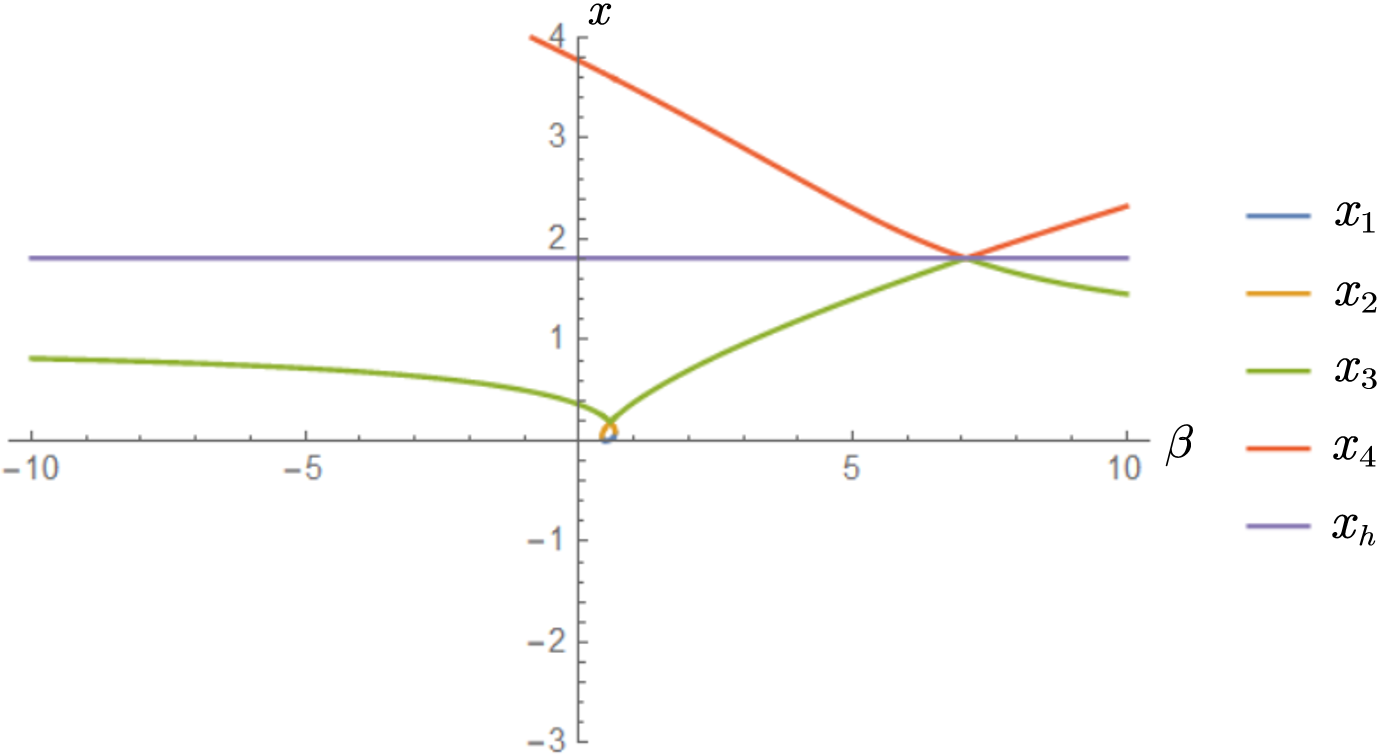} % 插入图片，[]中设置图片大小，{}中是图片文件名
    \caption{Radii of four orbits and event horizon radius as functions of $\beta$ with fixed parameters $u=0.5,w=0.3$.} % 最终文档中希望显示的图片标题
    \label{kngsol} % 用于文内引用的标签
\end{figure}
\begin{figure}[H] % H为当前位置，!htb为忽略美学标准，htbp为浮动图形
    \centering % 图片居中
    \includegraphics[width=0.35\textwidth]{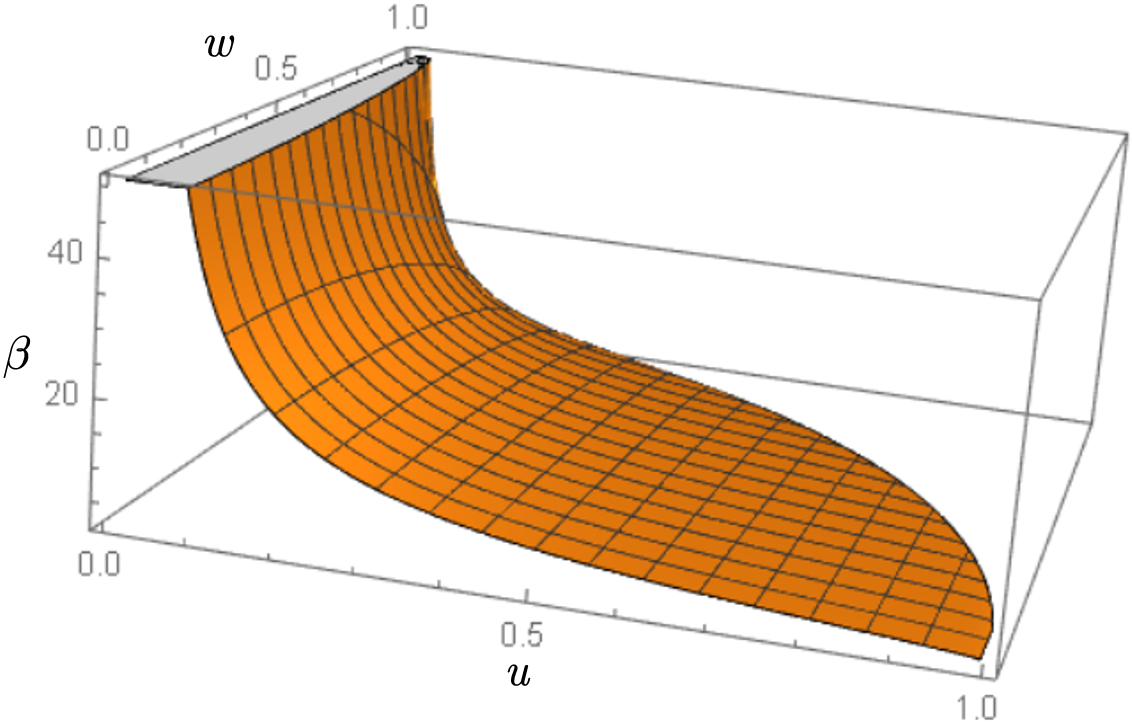} % 插入图片，[]中设置图片大小，{}中是图片文件名
    \caption{The no-orbit surface in $(u,w,\beta)$ space.} % 最终文档中希望显示的图片标题
    \label{kngc} % 用于文内引用的标签
\end{figure}

 % 结束环境
From Fig.\ref{kngsol}, we can also note that there exists a special value for $\beta$ where the orbits $x_3,x_4$ lies on the event horizon. There is no orbit outside the event horizon. In fact, this is a general result.   
It can be demonstrated that there exists a surface (dubbed no-orbit surface) in the \( (u, w, \beta) \) parameter space. When a point $(u,w,\beta)$ lies on this surface, there are no orbits outside the black hole's event horizon.  When \( (u, w, \beta) \) is not on the surface, there is always one orbit outside the event horizon. This surface can be determined by the condition \( x_3 = x_4 = x_h \), which is plotted in Fig.\ref{kngc}.
 % 结束环境

It should also be noticed that in this section (\( \gamma = 1 \)), the spherical orbits outside the event horizon are always unstable. It is easy to understand these results since a stable orbit should be in bound state $\gamma <1$.
 
\section{Radii of spherical timelike orbits with $\gamma\neq 1$}
In this section we focus on the radii of spherical timelike polar and equatorial orbits around Kerr-Newman black holes.

\subsection{Polar orbits}\label{pl}
For polar orbits, the angular momentum parameter of the particle vanishes, i.e. $\beta=0$. Then from Eq.\eqref{P5}, we can obtain the quintic equation satisfied by the radii of the polar orbits, which is
\begin{equation}\label{po-gn1}
	\begin{split}
	&\left(\gamma ^2-1\right) x^5+\left(4-3\gamma ^2\right) x^4\\
	&+\left(2\gamma ^2 u^2-2 u^2+2 \gamma ^2 w^2-2 w^2-4\right)x^3\\
	&+\left(-2 \gamma ^2 u^2+4 u^2+4w^2\right)x^2\\
	&+\left(\gamma ^2 u^4-u^4+2 \gamma^2 u^2 w^2-2 u^2 w^2-w^4\right)x\\
	&+\gamma ^2 u^4=0 .
	\end{split}
\end{equation}
Since there are no general analytic solutions to a quintic equation, we will conduct the analysis for several typical values of $\gamma$ by using numerical method.

\begin{figure}[H] % H为当前位置，!htb为忽略美学标准，htbp为浮动图形
    \centering % 图片居中
    \includegraphics[width=0.4\textwidth]{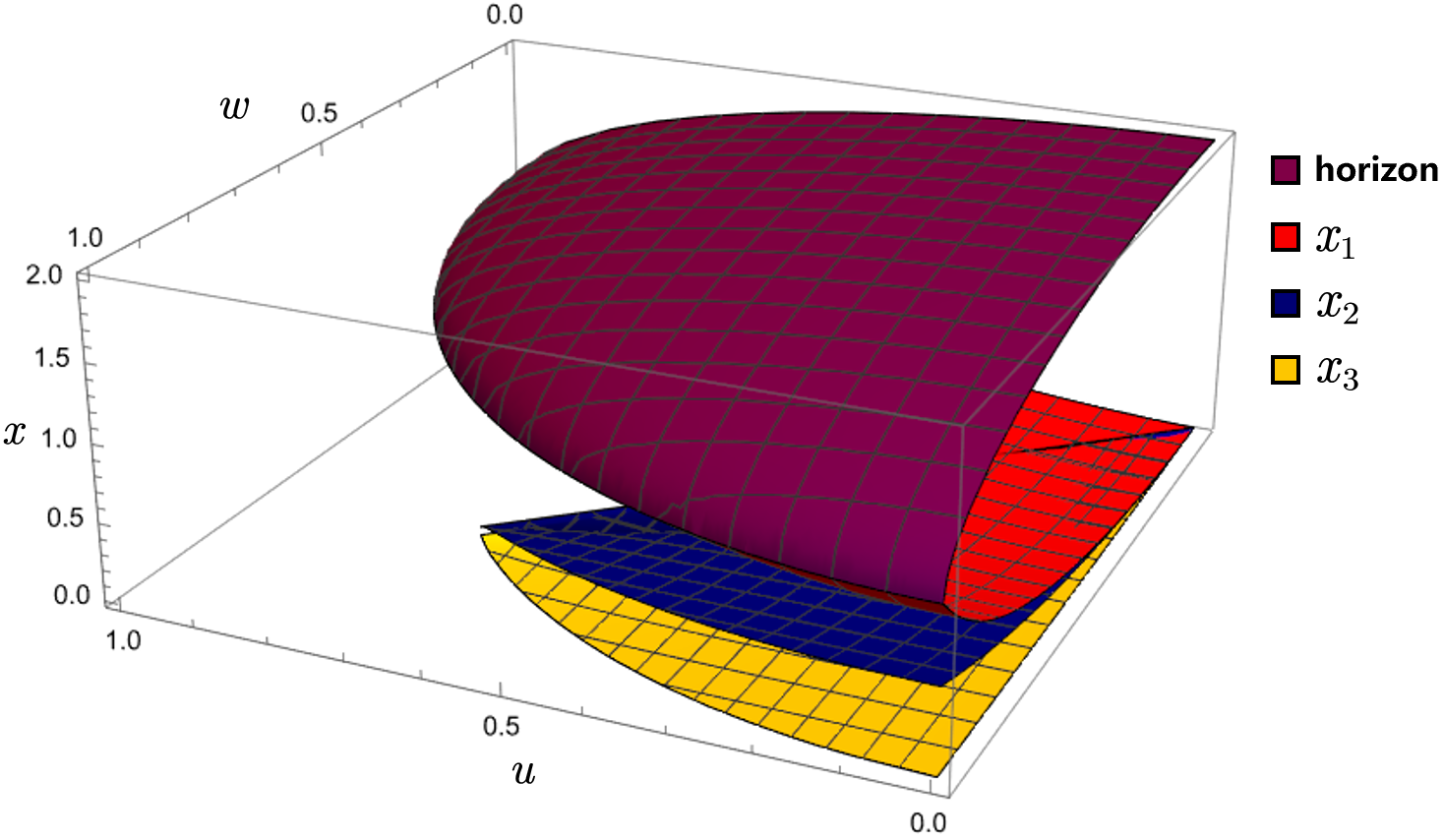} % 插入图片，[]中设置图片大小，{}中是图片文件名
    \caption{Event horizon radius and the radii of the three orbits are plotted as functions of $(u,w)$ when $\gamma=0.5$. No orbit is outside the event horizon.} % 最终文档中希望显示的图片标题
    \label{knpls5} % 用于文内引用的标签
\end{figure}

For \( 0 < \gamma < 1 \), we first choose \( \gamma = 0.5 \). It is found that for different values of $(u,w)$, there are one or three real roots to the Eq.\eqref{po-gn1}. We plot them and the event horizon radius in Fig.\ref{knpls5}. It is noticed that the three orbits are all inside the event horizon, and there is no orbit outside the event horizon. 
 % 结束环境

\begin{figure}[H] % H为当前位置，!htb为忽略美学标准，htbp为浮动图形
    \centering % 图片居中
    \includegraphics[width=0.4\textwidth]{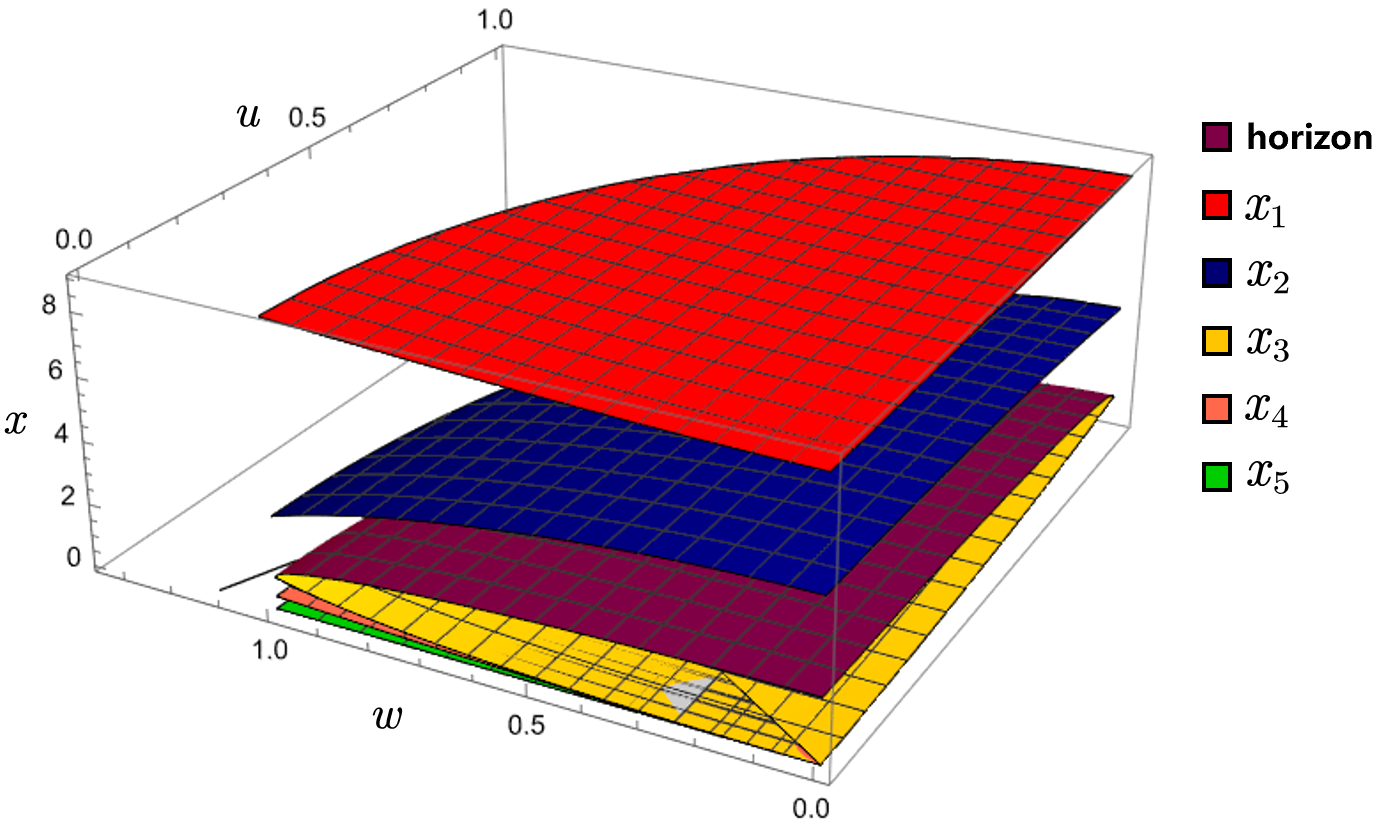} % 插入图片，[]中设置图片大小，{}中是图片文件名
    \caption{Event horizon radius and the radii of the three orbits are plotted as functions of $(u,w)$ when $\gamma=0.95$. Two orbits with radii $x_1,x_2$ are outside the event horizon.} % 最终文档中希望显示的图片标题
    \label{knpls95} % 用于文内引用的标签
\end{figure}

\begin{figure}[H] % H为当前位置，!htb为忽略美学标准，htbp为浮动图形
    \centering % 图片居中
    \includegraphics[width=0.4\textwidth]{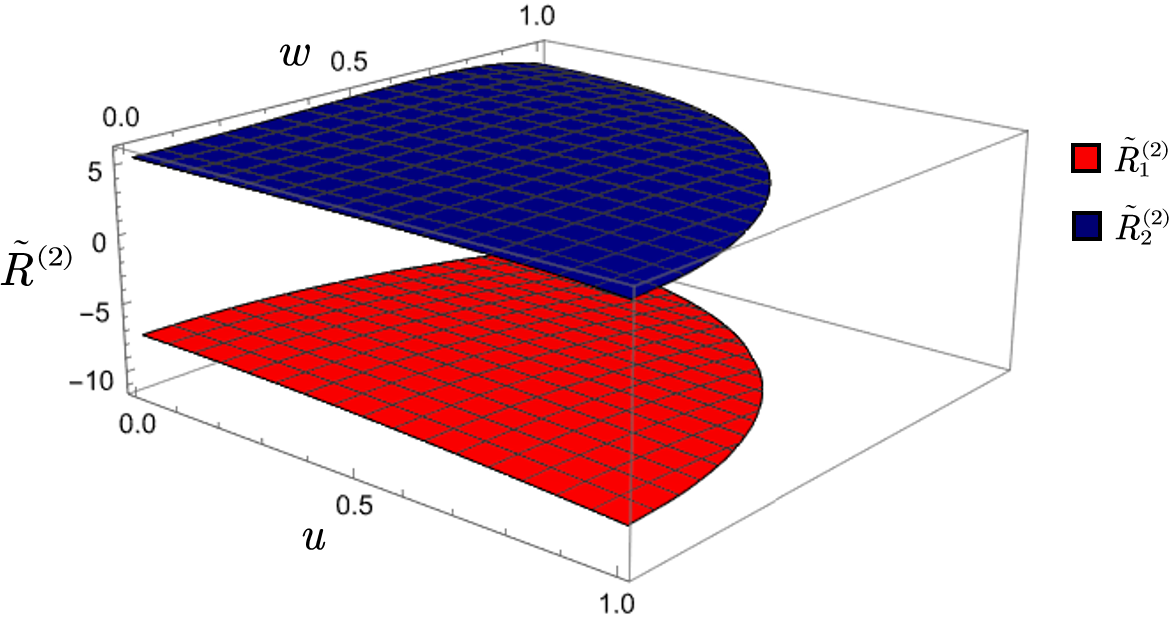} % 插入图片，[]中设置图片大小，{}中是图片文件名
    \caption{$\tilde{R}^{(2)}_i$ for the orbit with radius $x_i (i=1,2)$  when $\gamma=0.95$.} % 最终文档中希望显示的图片标题
    \label{knplrq95} % 用于文内引用的标签
\end{figure}

We then consider \( \gamma = 0.95 \). It is found that for different values of $(u,w)$, there are five or three real roots to the Eq.\eqref{po-gn1}. We plot them and the event horizon radius in Fig.\ref{knpls95}. It is interesting that there are always two orbits (with radii $x_1,x_2$) outside the event horizon.
 % 结束环境
In order to check the radial stability of the two orbits outside the event horizon, we plot \( \tilde{R}^{(2)}_i \) corresponding to these two orbits as functions of \( u \) and \( w \) in Fig.\ref{knplrq95}. We can see that the orbit with the smaller radius \( x_2 \) is unstable under radial perturbation, while the orbit with the larger radius \( x_1 \) is stable.
 % 结束环境

\begin{figure}[H] % H为当前位置，!htb为忽略美学标准，htbp为浮动图形
    \centering % 图片居中
    \includegraphics[width=0.4\textwidth]{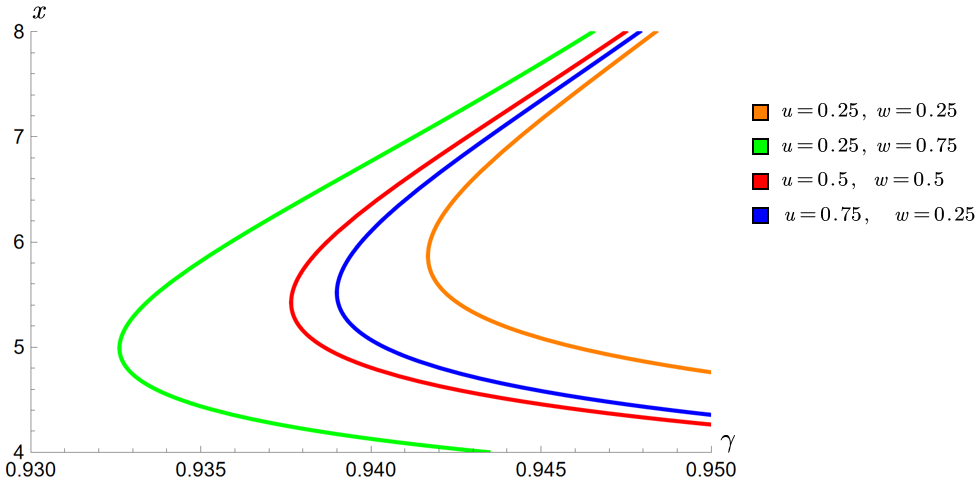} % 插入图片，[]中设置图片大小，{}中是图片文件名
    \caption{Radius $x$ of a spherical polar orbit as a function of particle energy $\gamma$ ($0 < \gamma < 1$) for different values of $u$ and $w$.} % 最终文档中希望显示的图片标题
    \label{splgax} % 用于文内引用的标签
\end{figure}

From the above discussion of the two cases with $\gamma=0.5,0.95$, we see that there are two or no spherical orbits outside the event horizon. 
In fact, when $0 < \gamma < 1$, there exists a boundary surface in the parameter space $(u, w, \gamma)$ that separates the parameter regions without spherical orbits and with two spherical orbits outside the event horizon. And when parameters are on this boundary surface,  there is only one spherical orbits outside the event horizon.
To illustrate this, we fix the values of $u$ and $w$ in Eq.\eqref{po-gn1}, and plot the radii of spherical orbits, which are greater than the event horizon radius, as functions of particle energy $\gamma$ in Fig.\ref{splgax}. 
 % 结束环境
From Fig.\ref{splgax}, we can see that for given values of $u$ and $w$, there exists a minimum value $\gamma_{\mathrm{min}}$. When $\gamma_{\mathrm{min}} < \gamma < 1$, there are two different values of $x$, indicating the existence of two orbits outside the event horizon (consistent with the case of $\gamma = 0.95$). The larger orbit is stable and the smaller one is unstable. On the other hand, when $0 < \gamma < \gamma_{\mathrm{min}}$, no orbit exists outside the event horizon (as the case of $\gamma = 0.5$). When $\gamma = \gamma_{\mathrm{min}}$, there exists only a single orbit outside the event horizon.

\begin{figure}[H] % H为当前位置，!htb为忽略美学标准，htbp为浮动图形
    \centering % 图片居中
    \includegraphics[width=0.4\textwidth]{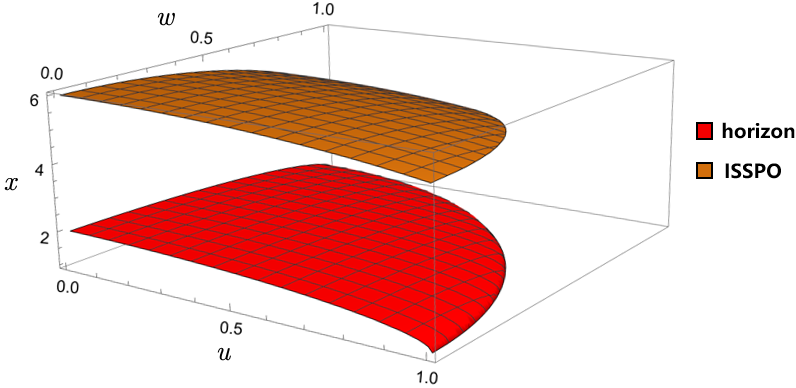} % 插入图片，[]中设置图片大小，{}中是图片文件名
    \caption{Radius $x$ of ISSPO is plotted as a function of $u$ and $w$.} % 最终文档中希望显示的图片标题
    \label{PISCO} % 用于文内引用的标签
\end{figure}

\begin{figure}[H] % H为当前位置，!htb为忽略美学标准，htbp为浮动图形
    \centering % 图片居中
    \includegraphics[width=0.4\textwidth]{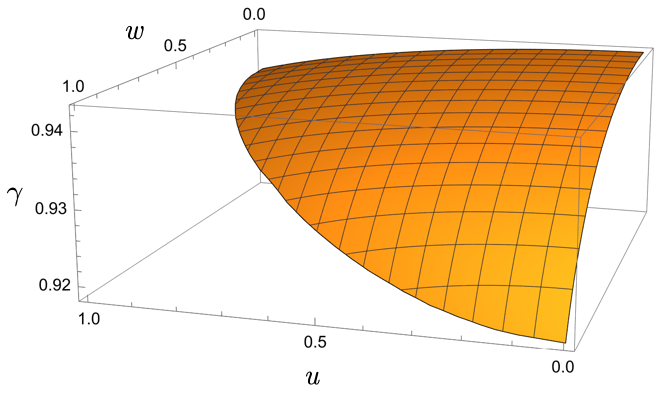} % 插入图片，[]中设置图片大小，{}中是图片文件名
    \caption{Boundary surface in the parameter space $(u, w, \gamma)$. For the existence of spherical polar orbits outside the event horizon, $\gamma$ need to be approximately larger than 0.92.} % 最终文档中希望显示的图片标题
    \label{PISCOgamma} % 用于文内引用的标签
\end{figure}

It will be interesting to analytically determine the boundary surface in the parameter space $(u, w, \gamma)$, which should be an equation involving $(u, w, \gamma)$. From the above discussion, we know that a point $(u_b, w_b, \gamma_b)$ on the boundary surface can also be interpreted as the energy $\gamma_b$ of a particle moving on the innermost stable spherical polar orbit (ISSPO) around a Kerr-Newman black hole with fixed parameters $(u_b,w_b)$. The radius of ISSPO can be determined by the spherical orbit condition, Eq.\eqref{r=0}, and an additional condition $\tilde{R}^{(2)}=0$, i.e.
\begin{equation}\label{r2=0}
    (8\gamma^2-8)x^3+(18-12\gamma^2)x^2-12x-4u^2\gamma^2+2(u^2+w^2)=0.
\end{equation}
From the above equation, we can obtain
\begin{equation}\label{gamma2}
    \gamma^2=\frac{u^2+w^2-6x+9x^2-4x^3}{2(u^2+3x^2-2x^3)}.
\end{equation}
Plugging the above equation \eqref{gamma2} and $\beta=0$ into Eq.\eqref{P5}, we finally get the equation satisfied by the radius of ISSPO
\begin{equation}
    \begin{split}
        &x^6-6x^5+(3u^2+9w^2)x^4+(4u^2-4w^2)x^3\\
        &+(3u^4-6u^2w^2)x^2-6u^4x+u^6+u^4w^2=0.
    \end{split}
\end{equation}
According to the above equation, the radius of ISSPO can be considered as a function of $u$ and $w$, which is plotted in Fig.\ref{PISCO}.
 % 结束环境
With Eqs.\eqref{po-gn1} and \eqref{r2=0}, by eliminating the higher order terms in $x$, we can obtain the following quadratic equation in $x$ satisfied by the radius of ISSPO,
\begin{equation}
    c_1x^2+c_2x+c_3=0\ ,
\end{equation}
where
\begin{equation}
    \begin{split}
        c_1=&\frac{-6 a_2 \gamma ^2+48 a_1^2 a_3 u^2+48 a_1^2 a_4 w^2+39}{64 a_1^2 \gamma ^2}\ ,\\
        c_2=&\frac{32 a_1^3 u^4+2 a_1 a_7 u^2-32 a_1^2 w^4+2 a_6 w^2-3 a_5}{32 a_1^2 \gamma ^2}\ ,\\
        c_3=&\frac{32 a_1^2 a_4 u^4+a_8 u^2+w^2 \left(a_5-32 a_1^2 w^2\right)}{64 a_1^2 \gamma ^2}\ ,\\
        a_1=&\gamma^2-1\ ,a_3=2\gamma^2-1\ ,\ a_4=4\gamma^2-1\ ,\\
        a_2=&55-86\gamma^2+36\gamma^4\ , \ a_5=23-56\gamma^2+36\gamma^4\ ,\\
        a_6=&55-109\gamma^2+54\gamma^4\ ,\\
        a_7=&32 a_1^2 w^2-12 \gamma ^4+68 \gamma ^2-55\ ,\\
        a_8=&64 a_1^3 w^2-72 \gamma ^6+148 \gamma ^4-102 \gamma ^2+23.
    \end{split}
\end{equation}
The solutions of the above quadratic equation are
\begin{equation}
    x_\pm=\frac{-c_2\pm\sqrt{c_2^2-4c_1c_3}}{2c_1}.
\end{equation}
We can numerically check that $x_+$ is outside the event horizon and is the radius of ISSPO. Plugging $x_+$ into Eq.\eqref{gamma2}, we can get the following equation for parameters
$(u, w, \gamma)$,  
\begin{equation}\label{bsurfpolar}
    \gamma^2=\frac{u^2+w^2-6x_++9x_+^2-4x_+^3}{2(u^2+3x_+^2-2x_+^3)}\ ,
\end{equation}
which determines the boundary surface in the parameter space $(u, w, \gamma)$ that separates the regions with and without orbits outside the event horizon. This boundary surface is numerically plotted in Fig.\ref{PISCOgamma}. When the parameters $(u, w, \gamma)$ lie above this surface in the parameter space, there are two spherical polar orbits outside the event horizon. When the parameters $(u, w, \gamma)$ lie on this surface, there is only one spherical polar orbit outside the event horizon.  When they lie below this surface, no spherical polar orbit exists outside the event horizon. 
 % 结束环境

\begin{figure}[H] % H为当前位置，!htb为忽略美学标准，htbp为浮动图形
    \centering % 图片居中
    \includegraphics[width=0.4\textwidth]{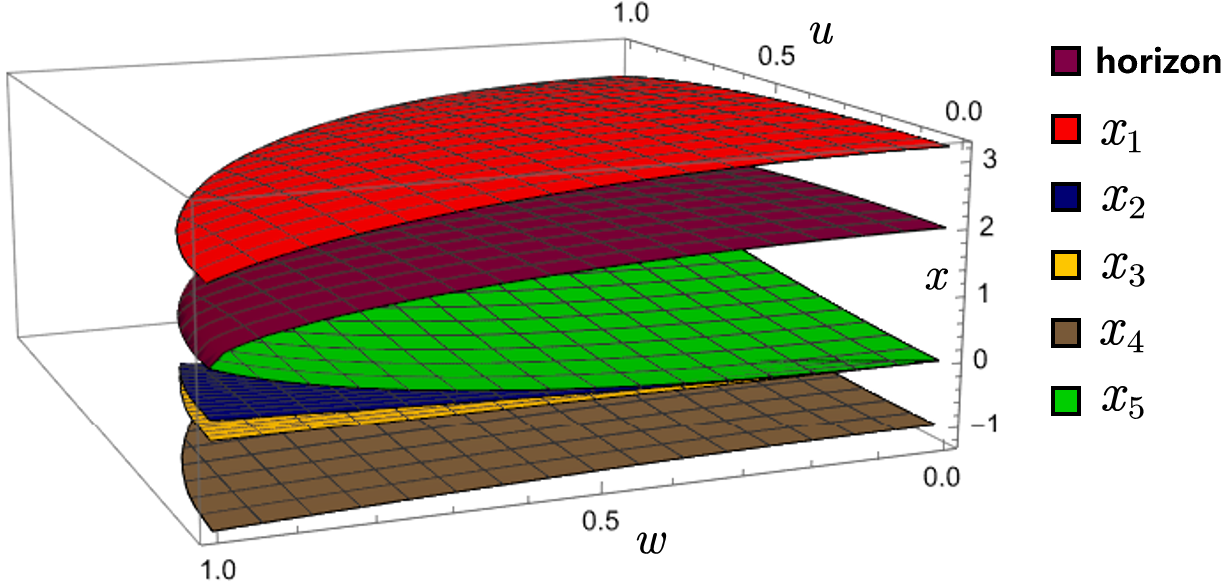} % 插入图片，[]中设置图片大小，{}中是图片文件名
    \caption{Event horizon radius and the radii of the five orbits are plotted as functions of $(u,w)$ when $\gamma=1.5$. One orbit with radius $x_1$ is outside the event horizon.} % 最终文档中希望显示的图片标题
    \label{knpll} % 用于文内引用的标签
\end{figure}

\begin{figure}[H] % H为当前位置，!htb为忽略美学标准，htbp为浮动图形
    \centering % 图片居中
    \includegraphics[width=0.4\textwidth]{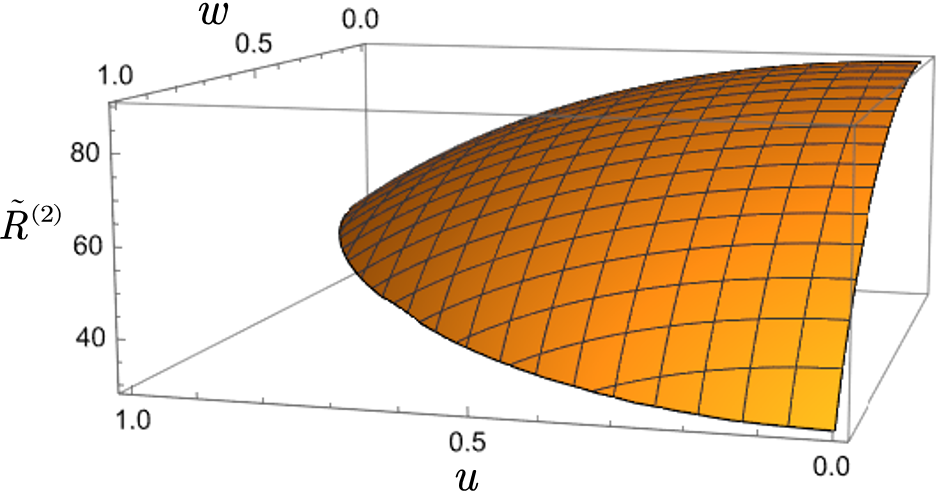} % 插入图片，[]中设置图片大小，{}中是图片文件名
    \caption{$\tilde{R}^{(2)}_1$ for the orbit with radius $x_1$ when $\gamma=1.5$.} % 最终文档中希望显示的图片标题
    \label{knpllrq} % 用于文内引用的标签
\end{figure}

For $\gamma>1$, we consider \( \gamma = 1.5 \). It is found that for different values of $(u,w)$ there are five or three real roots to Eq.\eqref{po-gn1}. We plot them and the event horizon radius as functions of $(u,w)$ in Fig.\ref{knpll}.
It is noticed that there is only one spherical orbit with radius $x_1$ outside the event horizon. 
We also plot the second derivative of the radial effective potential \( \tilde{R}^{(2)}_1 \) at $x_1$ as a function of \( u \) and \( w \) in Fig.\ref{knpllrq}. 
It is easy to see that \( \tilde{R}^{(2)}_1 > 0 \), indicating that the orbit with radius \( x_1 \) is radially unstable.

 % 结束环境

 % 结束环境

\subsection{Equatorial orbits}
For equatorial orbits, the Carter constant vanishes, i.e. the righthand sides of Eq.\eqref{c1} and Eq.\eqref{c2} are equal to $0$. The righthand side of Eq.\eqref{c1} can be considered as a quadratic equation in $\beta$, and with the dimensionless parameters defined in Eq.\eqref{diml}, we can solve the parameter $\beta$ and get
\bea\label{eqbeta}
&&\beta_{\pm}=\frac{- \gamma  u\pm\sqrt{ \gamma ^2 u^2- (x-1) K_1}}{x-1},\\
&&K_1=2\left(1- \gamma ^2\right) x^3-3 x^2\nonumber\\
&&~~~~+ \left(w^2-u^2(\gamma ^2 -1)\right)x-\gamma ^2 u^2.
\eea
$\beta_+>0$ corresponds to the prograde orbit, and $\beta_-<0$ corresponds to the retrograde orbit.

Plugging Eq.\eqref{eqbeta} into Eq.\eqref{c2}, we can get the equation satisfied by the radius of the equatorial spherical orbit,
\begin{equation}\label{eqeq}
    \begin{split}
     &\left(\gamma ^2-1\right) x^5+\left(5-4 \gamma ^2\right) x^4
     +\left((\gamma ^2-1) (2 w^2+3)-5\right)x^3 \\
       &+\left(-2 (\gamma ^2-1) (u^2+w^2)-u^2+4 (w^2+1)\right)x^2 \\
       &+\left(2 (\gamma ^2-1) u^2 w^2+u^2 (w^2-1)-w^2(w^2+4)\right)x\\
        &+(u^2+w^2) w^2\pm 2\gamma u(x-w^2) \sqrt{ \gamma ^2 u^2- (x-1) K_1}
  =0\ ,
    \end{split}
\end{equation}
where "$+$" corresponds to the prograde orbit and "$-$" corresponds to retrograde orbit.
There are no general analytic solutions to the above equation. 
We will consider several specific cases with \( \gamma =0.5, 0.97,1.5\) by using numerical method.

\begin{figure}[H]
    \centering
    \begin{subfigure}{0.4\textwidth}
        \includegraphics[width=\textwidth]{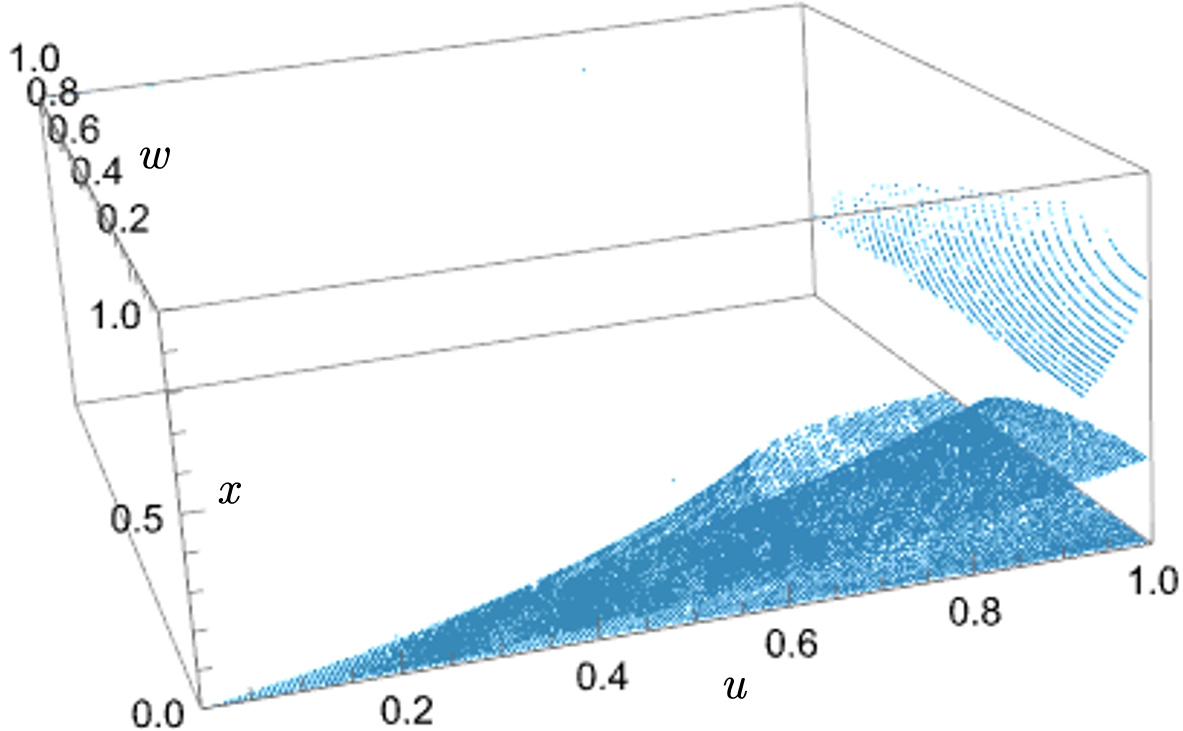}
        \caption{Prograde case.}
        \label{eqgm0.5}
    \end{subfigure}

   % \vspace{1em}  % 可选：增加上下图之间的空隙

    \begin{subfigure}{0.4\textwidth}
        \includegraphics[width=\textwidth]{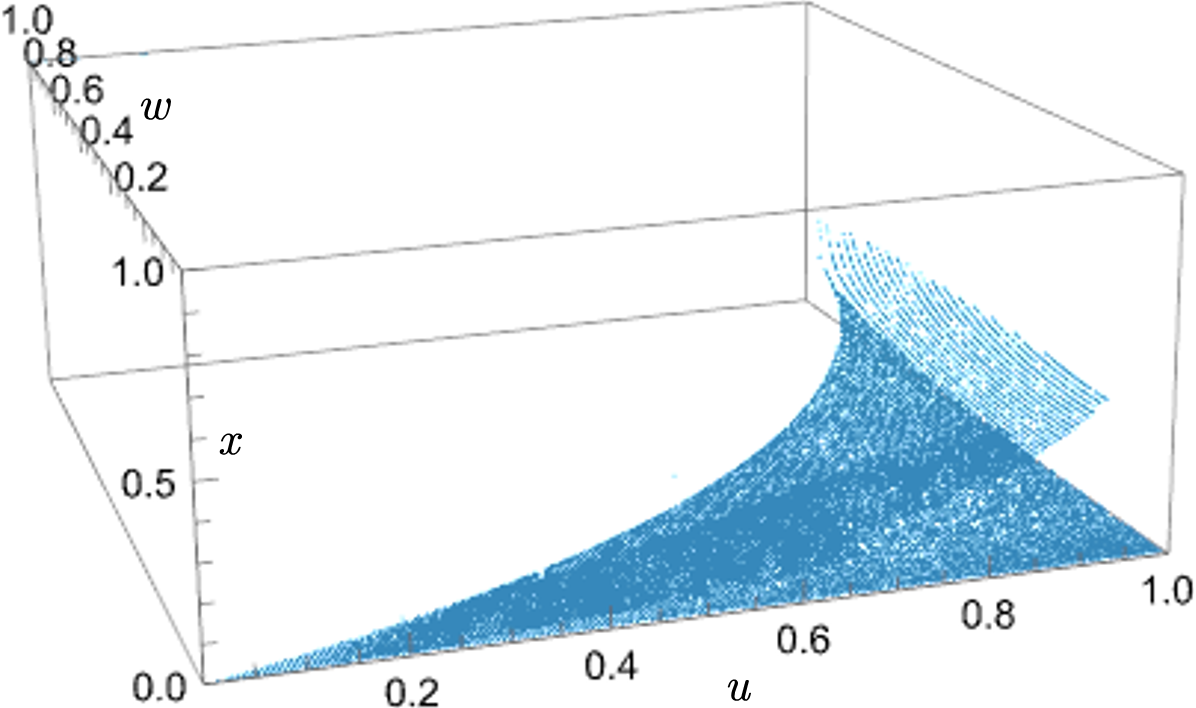}
        \caption{Retrograde case.}
        \label{eqgm0.5ni}
    \end{subfigure}

    \caption{Radii of orbits as functions of $(u,w)$ when $\gamma=0.5$. No orbit is outside the event horizon for a particle with energy $\gamma=0.5$. }
    \label{eqgm5}
\end{figure}

\begin{figure}[htbp]
    \centering
    \begin{subfigure}{0.4\textwidth}
        \includegraphics[width=\textwidth]{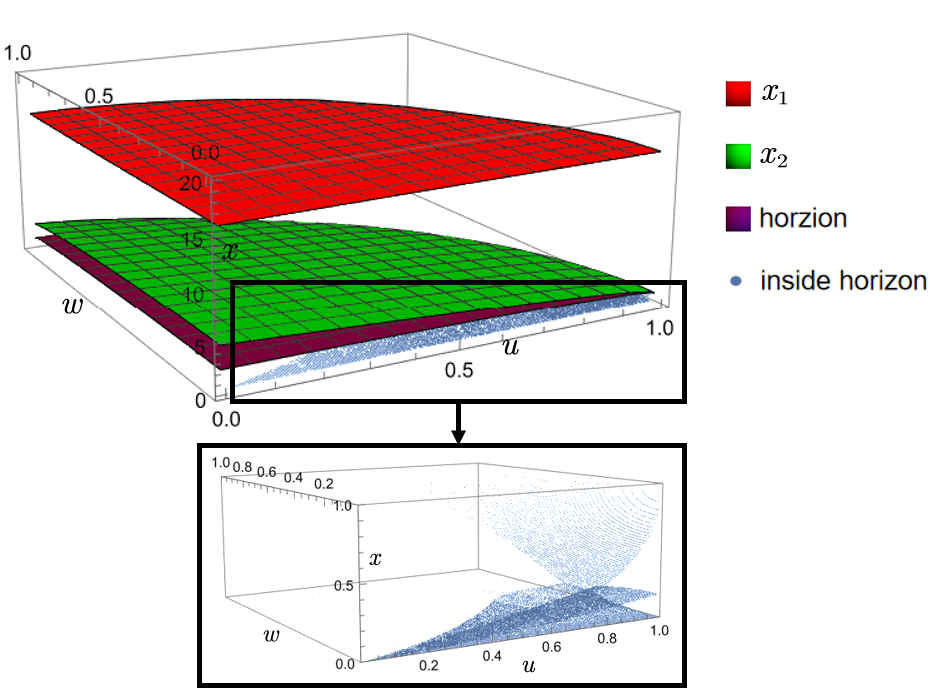}
        \caption{The prograde orbit when $\gamma=0.97$.}
        \label{eqgm095shun}
    \end{subfigure}

    \vspace{1em}  % 可选：增加上下图之间的空隙

    \begin{subfigure}{0.4\textwidth}
        \includegraphics[width=\textwidth]{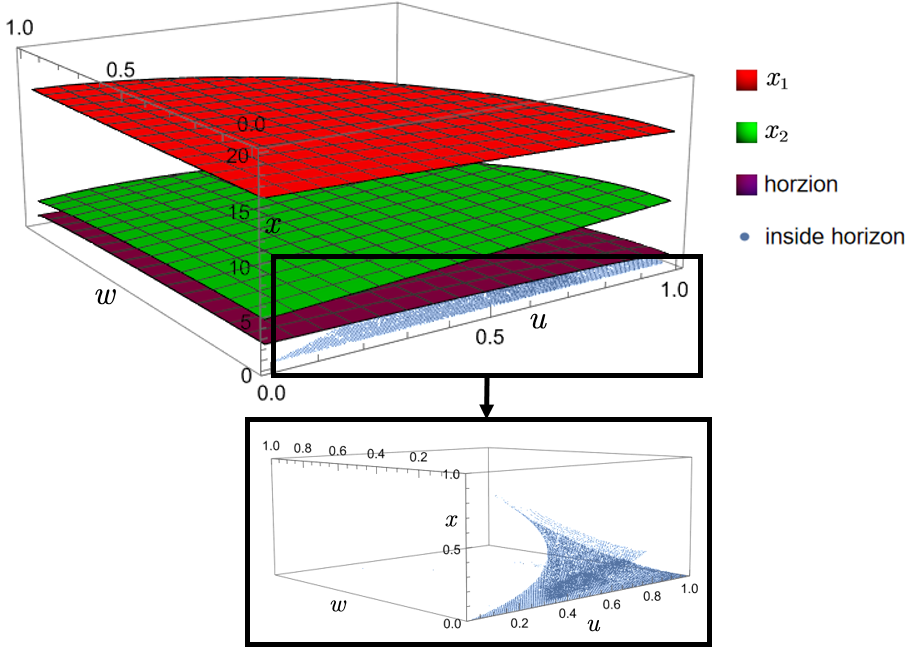}
        \caption{The retrograde orbit when $\gamma=0.97$.}
        \label{eqgm095ni}
    \end{subfigure}
    \captionsetup{justification=raggedright,singlelinecheck=false}
    \caption{Radii of orbits as functions of $(u,w)$ when $\gamma=0.97$. Two orbits are outside the event horizon for a particle with the same energy $\gamma=0.97$.}
    \label{eqgm095}
\end{figure}

\begin{figure}[htb]
    \centering
    \begin{subfigure}{0.35\textwidth}
        \includegraphics[width=\textwidth]{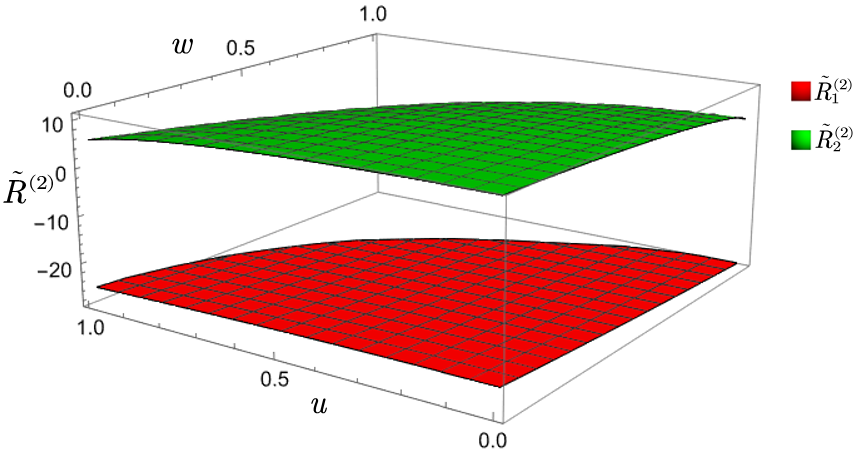}
        \caption{$\tilde{R}^{(2)}_i$ for the prograde orbits when $\gamma=0.97$.}
        \label{rqgm095shun}
    \end{subfigure}

    %\vspace{1em}  % 可选：增加上下图之间的空隙

    \begin{subfigure}{0.35\textwidth}
        \includegraphics[width=\textwidth]{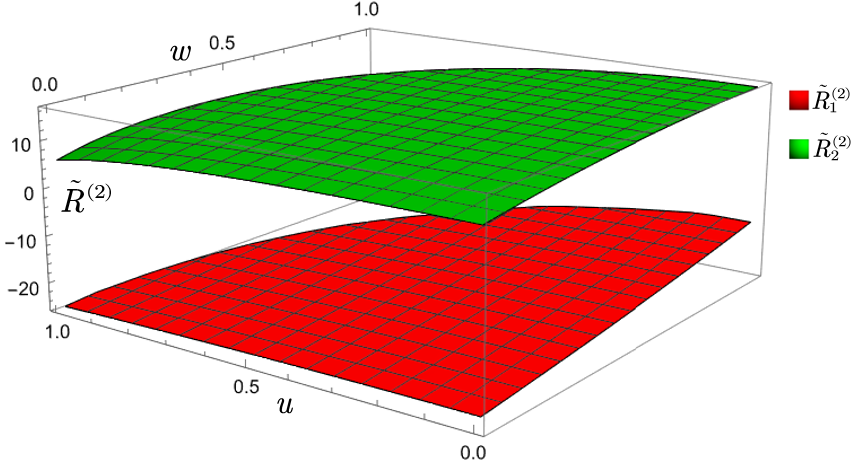}
        \caption{$\tilde{R}^{(2)}_i$ for the retrograde orbits when $\gamma=0.97$.}
        \label{rqgm095ni}
    \end{subfigure}
    \captionsetup{justification=raggedright,singlelinecheck=false}
    \caption{$\tilde{R}^{(2)}_i$ for the orbit with radius $x_i (i=1,2)$ when $\gamma=0.97$.}
    \label{rqgm095}
\end{figure}

For $0<\gamma<1$, we first consider $\gamma=0.5$. The real roots of Eq.\eqref{eqeq} are plotted as functions of $(u,w)$ in Fig.\ref{eqgm5}. It is found that all real roots of Eq.\eqref{eqeq} are smaller than $1$ and are inside the event horizon. The event horizon of Kerr-Newman black hole is located at $x_\mathrm{h}=1+\sqrt{1-u^2-w^2}$, which is greater than 1. So there is no timelike spherical orbit outside the event horizon when $\gamma=0.5$.
We then consider the case with $\gamma=0.97$. It is found that for different values of $(u,w)$ there are always two real roots for Eq.\eqref{eqeq} which are larger than the radius of the event horizon, and other roots (if exist) are inside the event horizon. We plot all roots as functions of $(u,w)$ in Fig.\ref{eqgm095}, and for the roots smaller than the event horizon radius, we just give the point plot with our numerical data.
In order to check the radial stability of the orbits outside the event horizon, we plot \( \tilde{R}^{(2)}_i \) corresponding to these orbits as functions of \( u \) and \( w \) in Fig.\ref{rqgm095}. We can see that the orbit with the smaller radius \( x_2 \) is unstable under radial perturbation, while the orbit with the larger radius \( x_1 \) is stable.
These two orbits are located on both sides of the inner most stable orbit (ISCO).

\begin{figure}[http]
    \centering
    \begin{subfigure}{0.4\textwidth}
        \includegraphics[width=\textwidth]{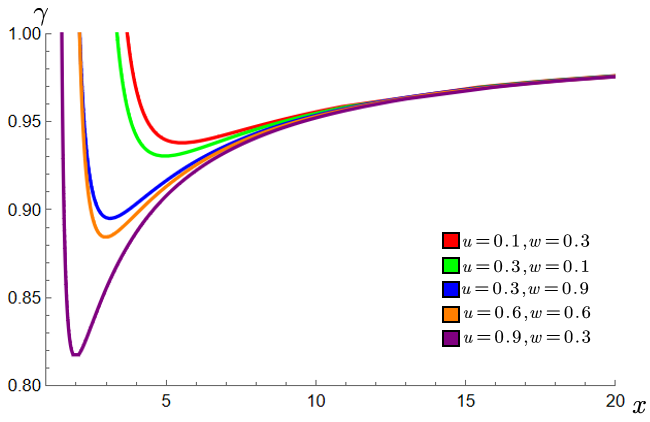}
        \caption{For prograde orbits, the relation between $\gamma$ and $x$  with different values of $u$ and $w$.}
        \label{eqxgs}
    \end{subfigure}

    %\vspace{1em}  % 可选：增加上下图之间的空隙

    \begin{subfigure}{0.4\textwidth}
        \includegraphics[width=\textwidth]{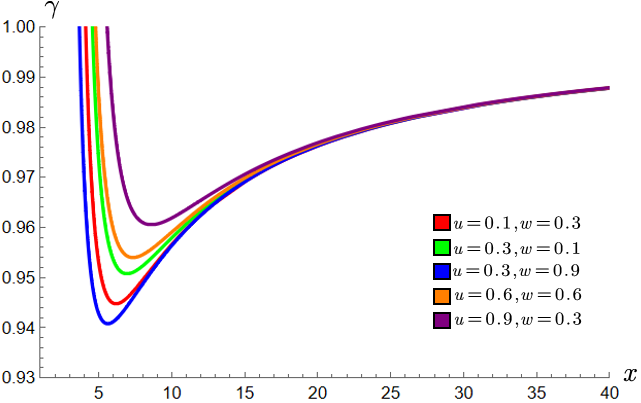}
        \caption{For retrograde orbits, the relation between $\gamma$ and $x$ with different values of $u$ and $w$.}
        \label{eqxgn}
    \end{subfigure}
    \captionsetup{justification=raggedright,singlelinecheck=false}
    \caption{The particle energy parameter $\gamma$ ($0 < \gamma < 1$) as a function of the radius of spherical orbit, $x$, for various values of $(u,w)$.}
    \label{eqxg}
\end{figure}

From the above discussion of the two cases with $\gamma=0.5,0.97$, we see that there may be two or no spherical equatorial timelike orbits outside the event horizon for particle with different energy parameter $\gamma$. 
In fact, when $0 < \gamma < 1$, there exists a boundary surface in the parameter space $(u, w, \gamma)$ that separates the parameter regions without spherical orbits and with two spherical orbits outside the event horizon. And when parameters are on this boundary surface,  there is only one spherical orbits outside the event horizon, which is just the ISCO.
To illustrate this, we first solve $\gamma$ and $\beta$ from the spherical orbit condition, Eq.\eqref{r=0}, with vanishing Carter constant and get 
\begin{equation}\label{gamabeta}
    \begin{split}
        \gamma_\pm&=\frac{\pm u \sqrt{x-w^2}+w^2+x^2-2 x}{x \sqrt{\pm 2 u \sqrt{x-w^2}+2 w^2+x^2-3 x}} \ ,\\
        \beta_\pm&=\frac{\pm\left(u^2+x^2\right) \sqrt{x-w^2}+u \left(w^2-2 x\right)}{x \sqrt{\pm2 u \sqrt{x-w^2}+2 w^2+x^2-3 x}}\ ,
    \end{split}
\end{equation}
 where "$+$" corresponds to the prograde orbit and "$-$" corresponds to retrograde orbit.   
 Then,  we fix the values of $u$ and $w$ in Eq.\eqref{gamabeta}, and plot the particle energy $\gamma$ as a function of radius $x$ of the spherical orbit which is outside the event horizon, see Fig.\ref{eqxg}. 

From Fig.\ref{eqxg}, we can see that for given values of KN black hole parameters $u$ and $w$, there exists a minimum value $\gamma_{\mathrm{min}}$. When $\gamma_{\mathrm{min}} < \gamma < 1$, there are two different values of $x$, indicating the existence of two orbits outside the event horizon (the $\gamma = 0.97$ case). The larger orbit is stable and the smaller one is unstable. On the other hand, when $0 < \gamma < \gamma_{\mathrm{min}}$, no orbit exists outside the event horizon (the $\gamma = 0.5$ case). When $\gamma = \gamma_{\mathrm{min}}$, there exists only a single orbit outside the event horizon, which is just the ISCO for black hole with the given parameter $u$ and $w$.

\begin{figure}[http]
    \centering
    \begin{subfigure}{0.4\textwidth}
        \includegraphics[width=\textwidth]{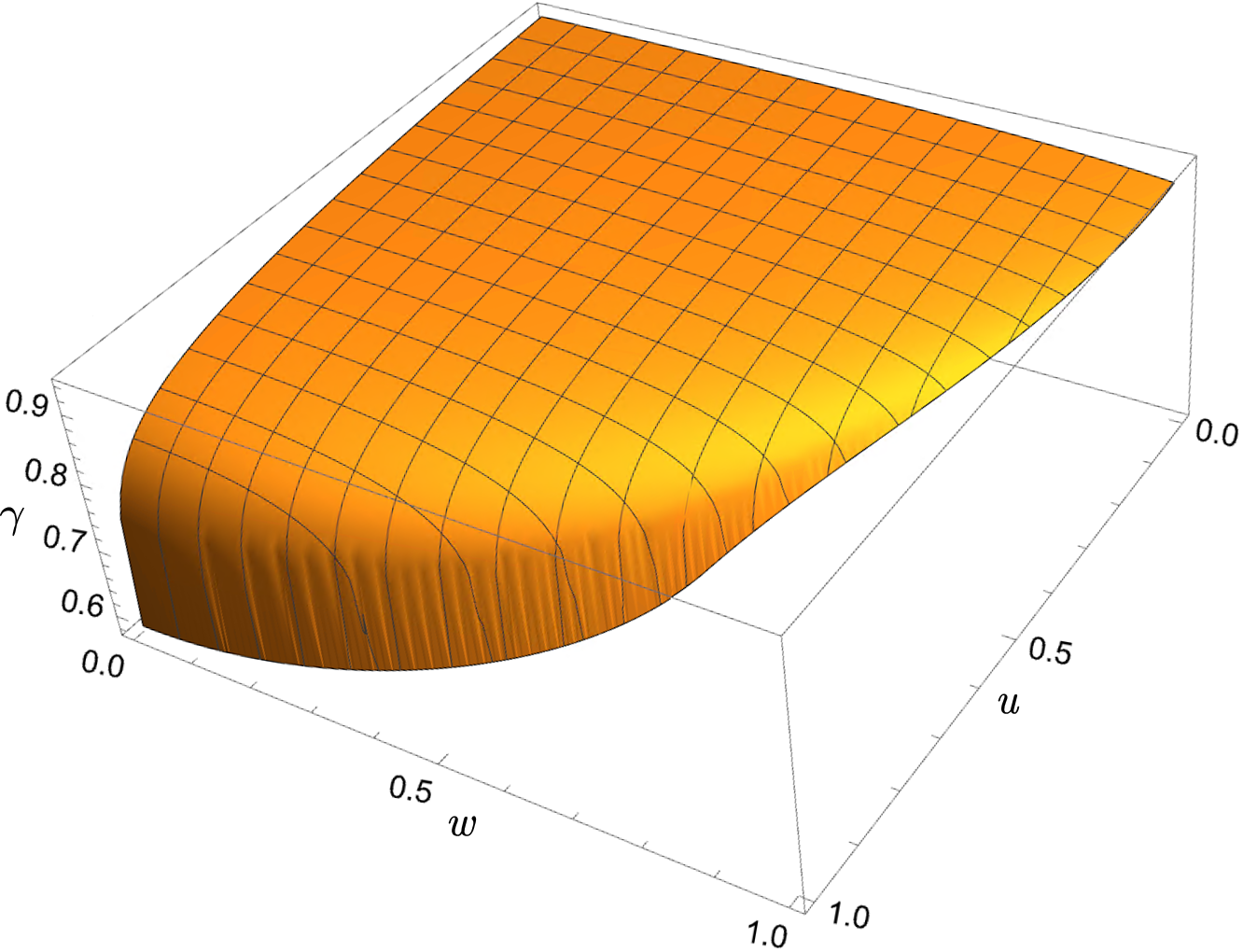}
        \caption{Prograde case.}
        \label{eqgamas}
    \end{subfigure}

    \vspace{1em}  % 可选：增加上下图之间的空隙

    \begin{subfigure}{0.4\textwidth}
        \includegraphics[width=\textwidth]{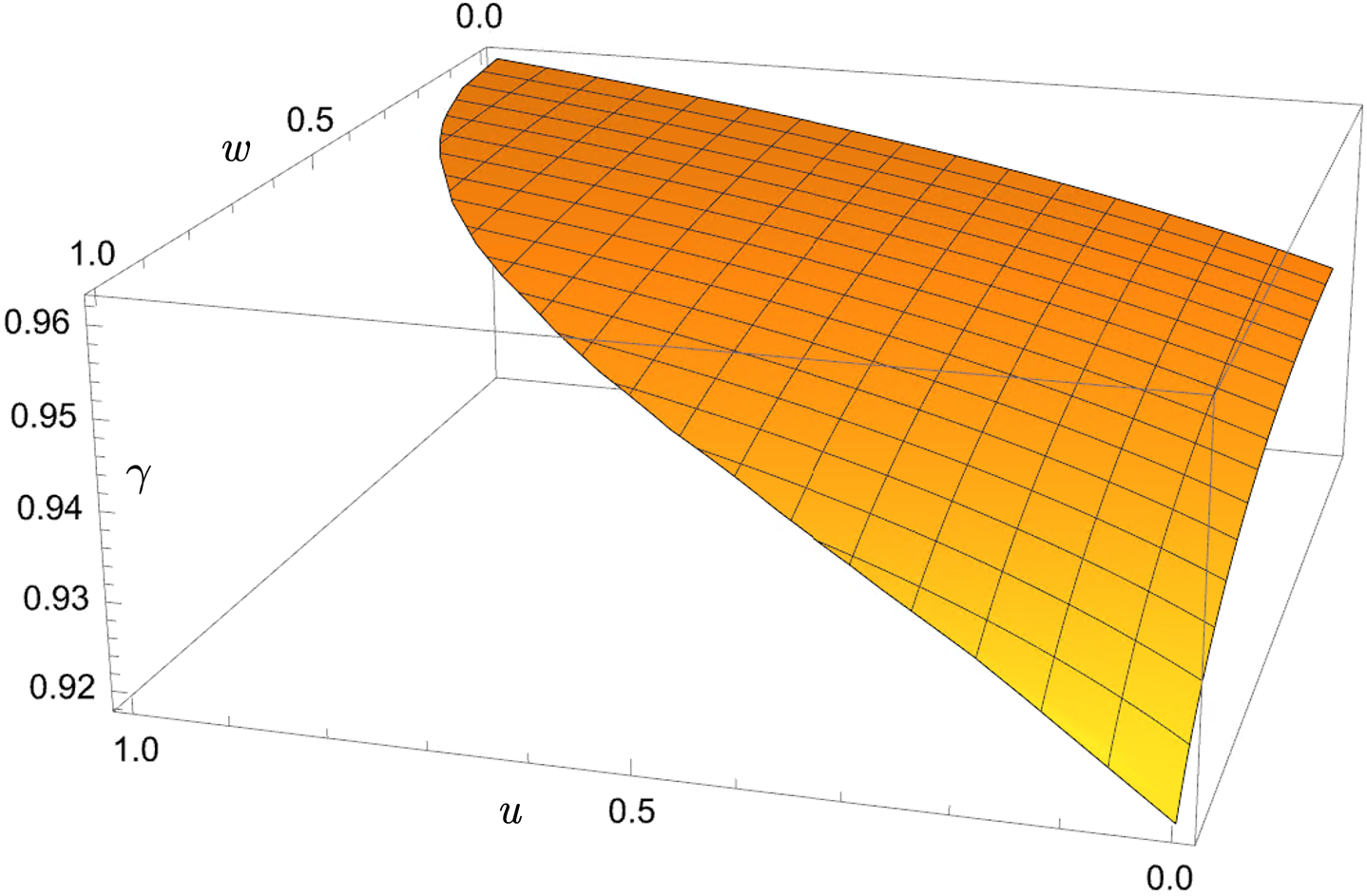}
        \caption{Retrograde case.}
        \label{eqgaman}
    \end{subfigure}
    \captionsetup{justification=raggedright,singlelinecheck=false}
    \caption{Boundary surfaces in the parameter space $(u, w, \gamma)$.}
    \label{eqgama}
\end{figure}

The ISCO radius is determined by the following three equations
\begin{equation}
    R(x; u, w,\beta,\gamma)=0,~~\frac{\mathrm{d}R}{\mathrm{d}x}=0,~~\frac{\mathrm{d}^2R}{\mathrm{d}x^2}=0,
\end{equation}
where $ x, u, w,\beta,\gamma$ are defined in \eqref{diml} and the Carter constant is 0 for equatorial orbits. 
Usually, for fixed black hole parameters $u,w$, one can just numerically calculate the ISCO radius, conserved energy and angular momentum of particles on ISCO from the above three equations. In order to obtain some analytical results, here we treat $x, \beta, u$ as functions of the particle energy $\gamma$ and black hole charge $w$. By solving the above three equations, the ISCO radius is
\bea\label{isco}
x_I&=&\frac{\sqrt{4 \left(\gamma ^2-1\right)^2 w^4+5 \left(\gamma ^2-1\right) w^2+1}}{3 \left(1-\gamma ^2\right)}\nonumber\\
&&+\frac{2(1- \gamma ^2) w^2+1}{3 \left(1-\gamma ^2\right)}.
\eea 
Note that although there are prograde and retrograde ISCOs, they both satisfy the above formula. The angular momentum parameters of the particle on prograde and retrograde ISCOs  are respectively 
\begin{equation}
    \begin{split}
\beta_I^p=&\sqrt{\left(\gamma ^2-1\right) (u^p)^2-w^2+6 x_I \left(\left(\gamma ^2-1\right) x_I+1\right)},\\
\beta_I^r=&-\sqrt{\left(\gamma ^2-1\right) (u^r)^2-w^2+6 x_I \left(\left(\gamma ^2-1\right) x_I+1\right)},
   \end{split}
\end{equation}
where $u^p$ and $u^r$ are the corresponding black hole rotation parameters, which are 
\bea
  u^p&=&\frac{1}{\gamma}(\beta_I^p- x_I\sqrt{3-4x_I+4x_I\gamma^2}),\\
  u^r&=&\frac{1}{\gamma}(\beta_I^r+ x_I\sqrt{3-4x_I+4x_I\gamma^2}).
\eea
The above formulas for $u^p$, $\beta_I^p$ and $x_I$ imply that $u^p$ and $\beta_I^p$ are functions of $\gamma$ and $w$. Similar result holds for $u^r$ and $\beta_I^r$. 
As we mentioned in previous paragraphs, there is a boundary surface in the $(u, w, \gamma)$ parameter space. For parameters on one side of this surface, there are two spherical orbits outside the event horizon, and for parameters on the other side of this surface, there is no spherical orbit outside the event horizon. For the prograde case, this boundary surface is determined by the following equation involving $(u, w, \gamma)$ 
\bea\label{bseqpr}
\gamma u&=&\sqrt{\left(\gamma ^2-1\right) u^2-w^2+6 x_I \left(\left(\gamma ^2-1\right) x_I+1\right)}\nonumber\\
&&- x_I\sqrt{3-4x_I+4x_I\gamma^2},
\eea 
where $x_I$ is determined by Eq.\eqref{isco}.
For the retrograde case, the boundary surface is determined by the following equation involving $(u, w, \gamma)$ 
\bea\label{bseqret}
\gamma u&=&-\sqrt{\left(\gamma ^2-1\right) u^2-w^2+6 x_I \left(\left(\gamma ^2-1\right) x_I+1\right)}\nonumber\\
&&+ x_I\sqrt{3-4x_I+4x_I\gamma^2},
\eea 
where $x_I$ is determined by Eq.\eqref{isco}.
We plot the two boundary surfaces in Fig.\ref{eqgama}. When the parameters $(u, w, \gamma)$ lie above the surfaces in the parameter space, there are two spherical orbits outside the event horizon. When $(u, w, \gamma)$ lie below the surfaces, no spherical orbit exists outside the event horizon. When  $(u, w, \gamma)$ lie on the surfaces, only one spherical orbit (i.e. ISCO) is outside the event horizon and now the value of $\gamma$ is the conserved energy of particles moving on the ISCO of black hole with parameters $(u,w)$.

\begin{figure}[http]
    \centering
    \begin{subfigure}{0.4\textwidth}
        \includegraphics[width=\textwidth]{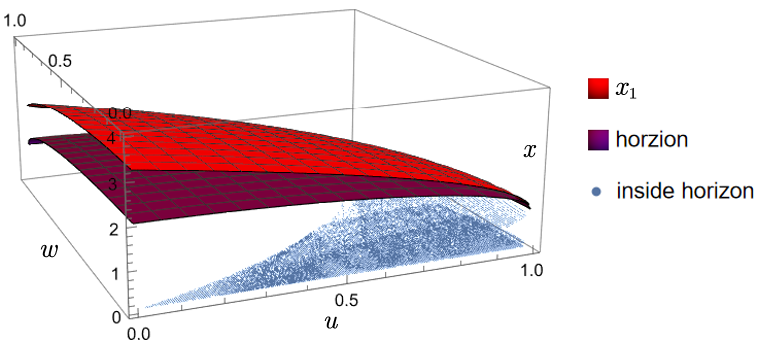}
        \caption{Prograde case.}
        \label{rqgm15s}
    \end{subfigure}

   % \vspace{1em}  % 可选：增加上下图之间的空隙

    \begin{subfigure}{0.4\textwidth}
        \includegraphics[width=\textwidth]{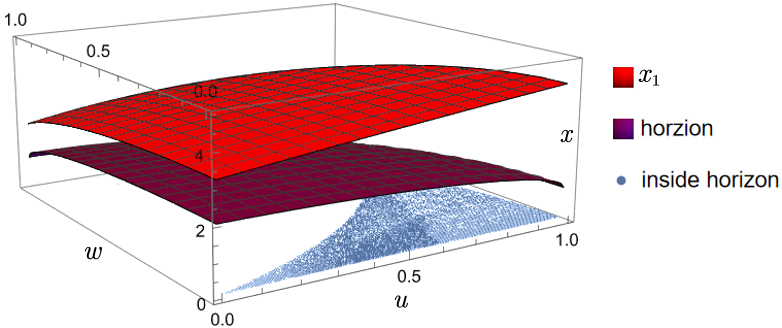}
        \caption{Retrograde case.}
        \label{eqgm15n}
    \end{subfigure}
    \captionsetup{justification=raggedright,singlelinecheck=false}
    \caption{Event horizon radius and the radii of the equatorial spherical orbits are plotted as functions of $(u,w)$ when $\gamma=1.5$. One orbit with radius $x_1$ is outside the event horizon.}
    \label{eqgm15}
\end{figure}

For $\gamma>1$, we consider \( \gamma = 1.5 \). It is found that for different values of $(u,w)$ there is always one root of Eq.\eqref{eqeq} larger than the radius of event horizon, and other roots are smaller than it. We plot all roots as functions of $(u,w)$ in Fig.\ref{eqgm15}. For the roots smaller than the event horizon radius, we just show the numerical data points.

\begin{figure}[http]
    \centering
    \begin{subfigure}{0.4\textwidth}
        \includegraphics[width=\textwidth]{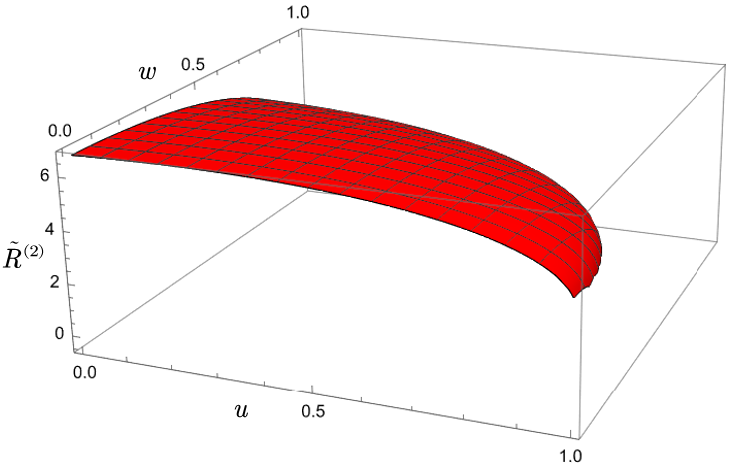}
        \caption{Prograde case.}
        \label{rqgm15s}
    \end{subfigure}

    \vspace{1em}  % 可选：增加上下图之间的空隙

    \begin{subfigure}{0.4\textwidth}
        \includegraphics[width=\textwidth]{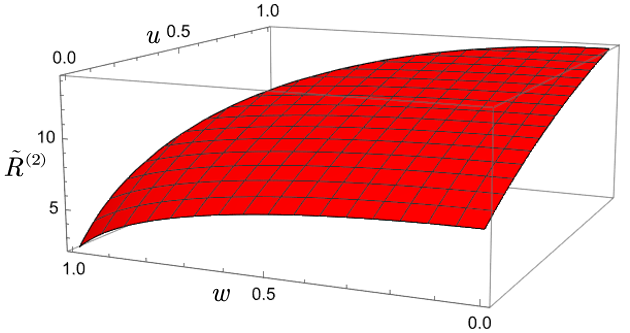}
        \caption{Retrograde case.}
        \label{rqgm15n}
    \end{subfigure}
    \captionsetup{justification=raggedright,singlelinecheck=false}
    \caption{$\tilde{R}^{(2)}_1$ for the orbit with radius $x_1$ when $\gamma=1.5$.}
    \label{rqgm15}
\end{figure}

We also plot the second derivative of the radial effective potential \( \tilde{R}^{(2)}_1 \) at $x_1$ as a function of \( u, w \) for prograde and retrograde cases in Fig.\ref{rqgm15}. 
It is easy to see that \( \tilde{R}^{(2)}_1 > 0 \), which indicates that the orbit with radius \( x_1 \) is radially unstable.

\section{conclusion}
In this paper, we analytically and numerically study the spherical geodesics of a neutral massive particle around a Kerr-Newman black holes. Starting from the radial geodesics equation of a massive particle in Kerr-Newman spacetime and the two spherical orbit conditions, we derive a quintic equation satisfied by the radii of the spherical orbits, which depends on four parameters, i.e. the rotation parameter $u$ and charge parameter $w$ of the black hole, the conserved angular momentum parameter $\beta$ and conserved energy $\gamma$ of the particle. We mainly focus on the analytical formula for the radii of spherical orbits, the number of orbits outside the horizon and the stability of the orbits.
According to the value of the conserved energy $\gamma$ of the particle, we discuss three cases with $\gamma=1$, $0<\gamma<1$ and $\gamma>1$. The main results we have found are:
\begin{itemize}
    \item Spherical polar orbits ($\beta=0$). For the $\gamma=1$ case, which we call critical case, there are four or two solutions to the corresponding equation \eqref{pp4}, but only one unstable orbit is outside the event horizon. The exact analytical formula for the radii are provided as functions of $(u, w)$. We also find the critical curve which separates two regions in $(u,w)$-plane corresponding to different number of solutions. For the $0<\gamma<1$ case, it is found that when $\gamma\lesssim 0.92$, there is no spherical orbit outside the event horizon, while for $\gamma\gtrsim 0.92$, there may be two spherical orbits outside the event horizon, and the orbit with a smaller radius is unstable and the one with larger radius is stable. There is a boundary surface in the parameter space $(u, w, \gamma)$. When parameters are chosen on this surface, there is only one spherical orbit outside the event horizon, which is the innermost stable spherical polar orbit (ISSPO). The equation to determine this boundary surface is given in Eq.\eqref{bsurfpolar}. For the $\gamma>1$ case, there is only one unstable orbit outside the event horizon. 
    \item Spherical equatorial orbits ($C=0$). For the $\gamma=1$ case, it is found that there are two classes of unstable orbits outside the event horizon in special regions in $(u,\beta)$-plane, shown in Fig.\eqref{kneqpara}. One class is the prograde ($\beta>0$) and the other is retrograde ($\beta<0$). Here, we choose $(u,\beta)$ as independent variables to obtain analytical formula for the radii of spherical equatorial orbits which are solutions of the quartic equation \eqref{peqc}. For the $0<\gamma<1$ case,   
        we consider two specific examples with $\gamma=0.5, 0.97$. The equation satisfied by the radii of the orbits is Eq.\eqref{eqeq}. When $\gamma=0.5$, it is found that all roots are smaller than the event horizon radius and no spherical orbit is outside the event horizon. When $\gamma=0.97$, it is found that there are two orbits outside the event horizon in each rotating case (prograde case and retrograde case), these two orbits are located on both sides of the ISCO, and the one with smaller radius is unstable while the other one is stable. In each rotating case, there is a boundary surface in $(u, w, \gamma)$ space that divides the space into two regions: one with two orbits outside the event horizon and the other with no orbit outside the event horizon. The equations to determine the boundary surfaces are given in Eq.\eqref{bseqpr} and Eq.\eqref{bseqret}, respectively. For parameters on the boundary surface in each rotating case, we obtain the corresponding ISCOs. An analytical formula for the radii of the ISCOs are obtained in Eq.\eqref{isco} by choosing $(w, \gamma)$ as independent parameters. This formula holds for both prograde and retrograde ISCOs.            
        For the $\gamma>1$ case, there is only one unstable orbit outside the event horizon.         
    \item General orbits. We just focus on the $\gamma=1$ case. It is found that there is a surface in $(u,w,\beta)$ space which is shown in Fig.\eqref{kngc}. When parameters $(u,w,\beta)$ lies on this surface, there is no orbit outside the event horizon, otherwise there is always one prograde or retrograde orbit outside the event horizon.
\end{itemize}

Here we mainly focus on several special classes of spherical timelike orbits around a Kerr-Newman black hole, and obtain analytical formulas for radii of these orbits. For the most general spherical timelike orbits, we can only use the numerical method to perform similar study. 
In this study, the angular motion of the particle is ignored. It will be interesting to take it into account and conduct further research on the motion of massive particles around 
a Kerr-Newman black hole.

\section*{Acknowledgments}
This work is partially supported by Guangdong Major Project of Basic and Applied Basic Research (No.2020B0301030008).

\section*{Appendix}\label{appendix}
For a quartic equation of $x$ with the following form
\begin{equation}
    a x^4+b x^3+c x^2+d x+e=0\ , 
\end{equation}   
 the general formulas of its solutions are
\begin{equation}
    \begin{split}
        x_1&=-\frac{A}{4}-\frac{1}{2} \sqrt{2 F-G-H-\frac{I }{4J}}-\frac{J}{2}\ ,\\
        x_2&=-\frac{A}{4}+\frac{1}{2} \sqrt{2 F-G-H-\frac{I }{4J}}-\frac{J}{2}\ ,\\
        x_3&=-\frac{A}{4}-\frac{1}{2} \sqrt{2 F-G-H-\frac{I }{4J}}+\frac{J}{2}\ ,\\
        x_4&=-\frac{A}{4}+\frac{1}{2} \sqrt{2 F-G-H-\frac{I }{4J}}+\frac{J}{2}\ ,
    \end{split}
\end{equation}
where
\begin{equation}
    \begin{split}
		A&=\frac{b}{a},\ B=\frac{c}{a},\ C=12 e a-3 b d+c^2,\\
		K&=\sqrt[3]{\sqrt{D^2-4C^3}+D}\ ,\  H=\frac{K}{3 \sqrt[3]{2} a},\\
        D&=-72 e a c+27 a d^2-9 b c d+27 e b^2+2 c^3\ ,\\
        E&=\frac{d}{a},\  F=\frac{A^2}{4}-\frac{2 B}{3},\  G=\frac{\sqrt[3]{2} C}{3 aK},\ ,\\
        I&=4 A B-A^3-8 E,\  J=\sqrt{F+G+H}\ .\\
    \end{split}
\end{equation}

\bibliography{refKNtime}
\end{document}